**Monitoring of Fluid Transport in Low Temperature Water Electrolyzers and Fuel Cells: Emerging Technologies and Future Prospects**

Zehua Dou [1, 5], Laura Tropf [3], Tobias Lappan [2], Hannes Rox [2], Xuegeng Yang [2], Lars Buettner [1], David Weik [1], Harry Hoster [3], Kerstin Eckert [2, 4, 5], Juergen Czarske [1]

[1] Faculty of Electrical and Computer Engineering, Laboratory of Measurement and Sensor System Technique, Technische Universität Dresden, Helmholtzstraße 18, 01069 Dresden, Germany

[2] Institute of Fluid Dynamics, Helmholtz-Zentrum Dresden-Rossendorf, Bautzner Landstraße 400, 01328Dresden, Germany

[3] Hydrogen and Fuel Cell Centre, ZBT GmbH, 47057 Duisburg, Germany

[4] Institute of Process Engineering and Environmental Technology, Technische Universität Dresden, Helmholtzstraße 14, 01069 Dresden, Germany

[5] Hydrogen Lab, School of Engineering, Technische Universität Dresden, Helmholtzstraße 14, 01062 Dresden, Germany

## Abstract

Low temperature water electrolyzers (LTWEs) and low temperature hydrogen fuel cells (LTFCs) present a promising technological strategy for the productions and usages of green hydrogen energy towards a net-zero world. However, the interactions of gas/liquid (fluid) transport and the intrinsic reaction kinetics in LTWEs/LTFCs present one of the key hurdles hindering high production rate and high energy conversion efficiency. Addressing these limitations requires analytical tools that are capable of resolving fluid transport across the heterogeneous, multiscale structures of operating LTWE and LTFC systems. This review provides a comprehensive overview of recent advancements in measurement technologies for investigating fluid transport. We first outline the technical requirements of such analytical systems, and assess the capabilities and limitations of established optical, X-rays and neutrons based imaging systems. We emphasis on emerging strategies that utilize integrated miniaturized sensors, ultrasound, and other alternative physical principles to achieve *operando*, high-resolution, and scalable measurements towards applications at device and system levels. Finally, we outline future directions in this highly interdisciplinary field, emphasizing the importance of next-generation sensing concepts to overcome the fluid transport hurdle, towards accelerating the deployment of green hydrogen technologies.

## 1. Introduction

The UN Secretary General, António Guterres, declared in July 2023 that "the era of global warming has ended and ***the era of global boiling has arrived***", after July was confirmed as the hottest month ever recorded.[1] Urgent energy transitions are needed to achieve the decarbonization for mitigating the climate change. Among emerging solutions, green hydrogen energy is widely recognized as a promising energy carrier, offering zero greenhouse gas emission [2], renewability and abundance [3], the highest specific power among all fuels [4], and transportability.

Up to date, hydrogen ($H_2$) can be produced in six different ways, as shown in **Fig. 1** (a), categorized by color codes based on their reactants, reaction conditions, by-products, and environmental impact. Among these methods, the green $H_2$ produced by water electrolysis using renewable electricity e.g., solar, wind turbines and marine energy is the only pathway offering long-term sustainability.[5,6] In turn, hydrogen fuel cells (FCs) can reconvert $H_2$ and $O_2$ into electricity and heat without carbon emissions, enabling applications from power supplies for spacecraft and aviation [7,8] to heavy-duty vehicles and stationary storage.[9,10]

Despite its potential, yet green hydrogen remains far from cost-competitive. In 2024, the green $H_2$ production accounted for less than 1 % of the total production, with only 2% used for power generation, while the cost was fivefold higher than fossil fuels', as illustrated in **Fig. 1** (b) & (c).[11] To achieve net zero emissions, the International Energy Agency (IEA) projects that hydrogen must supply 30% of energy demand [11], and the U.S. Department of Energy (DoE) targets a production cost of 1 USD/kg by 2031.[12] Achieving these goals requires technical breakthroughs in the designs, operation schemes and system integrations of commercial water electrolyzers (WEs) and fuel cells (FCs), particularly to improve the current densities, energy conversion efficiencies, and system durability.[13–15]

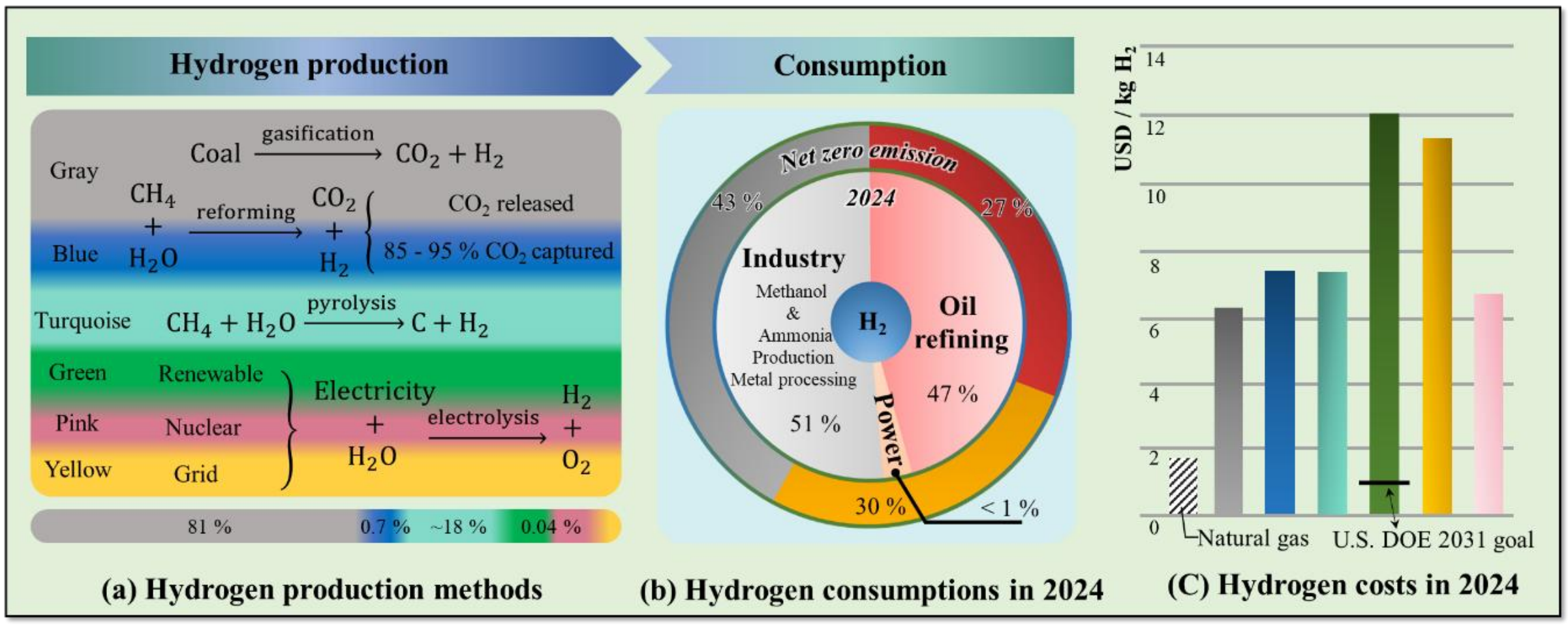

**Fig. 1. A non-exhaustive overview and a snapshot on the hydrogen energy.** (a) The different hydrogen production methods. (b) The hydrogen consumptions in 2024. (c) The cost of hydrogen in 2024.

As reciprocal devices, WEs and FCs often possess similar architectures and face comparable operational challenges. As shown in **Fig. 2** (a), WEs use externally supplied electricity to split water into $H_2$ and $O_2$, while the reverse reactions take place spontaneously in FCs, converting $H_2$ and $O_2$ to electricity and water. Based on the operating temperatures, state-of-the-art WE and FC technologies can roughly be classified into low temperature (LT, < 100 °C) and high temperature (HT, > 100 °C) schemes. High-temperature (HT) technologies e.g., solid oxide electrolyzers and fuel cells, drastically reduce the activation energy, offering enhanced efficiency and higher current densities.[16–19] These HT systems are however prone to degradation under high thermal stresses, limiting their long-term reliabilities and cost competitiveness.[20,21]

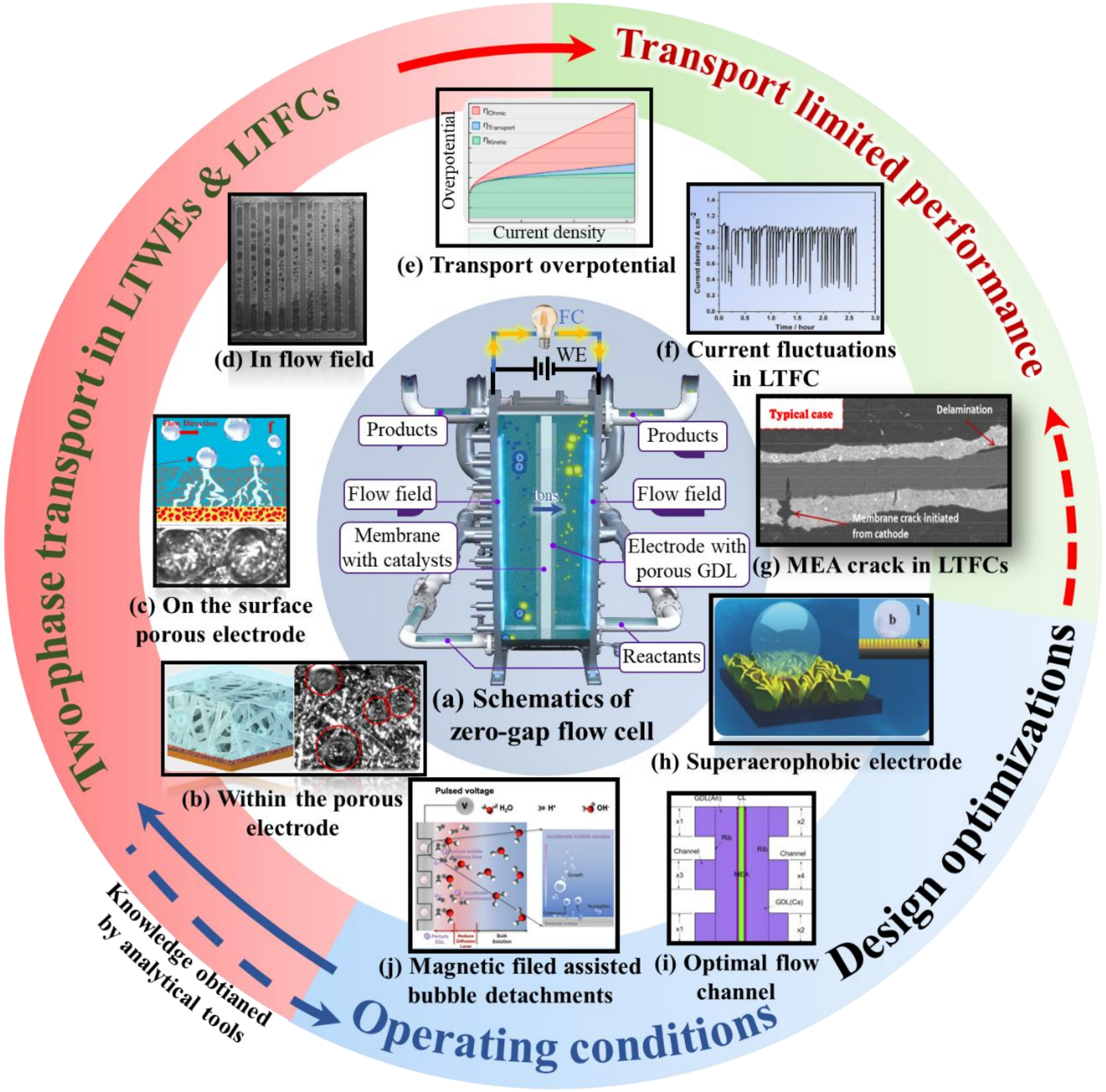

**Fig. 2. An overview of the complex interplays among designs, operating schemes, bubble and droplet dynamics in LTWEs and LTFCs, as well as their performance**. (a) The state-of-the-art zero-gap structure and working process of LTWEs and LTFCs. Image adapted from [22]. (b) – (d) Bubble and water evolutions in LTWEs & LTFCs: (b) bubbles/droplets confined in the pores of the GDL, images adapted from [23], (c) bubbles/droplets adhere to the surface of electrode, images adapted from [24], and (d) bubbles/droplets accumulated in the flow channels, image adapted from [25]. (e) – (g) The gas/liquid transport behaviours directly influence the performance of LTWEs and LTFCs: (e) the overpotential that decreases the voltage efficiency of a LTWE [26], (f) water accumulation in a PEMFC leads to its output current fluctuations when operated at the constant voltage mode [27], and (g) Membrane electrode assembly (MEA) degradations in PEMFC [28]. (h) – (i) Optimizations of designs and operating schemes for overcoming the transport limitations and improving the of LTWEs and LTFCs: (h) superaerophobic electrode for facilitating bubble removal in LTWEs [29], (i) optimal geometry of flow channels [30], and (j) pulse electrolysis scheme facilitating bubble removal in LTWEs [31].

In contrast, the LT technologies, based on alkaline [32], proton exchange membranes (PEMs) and anion exchange membranes (AEMs) [33–36] pose less challenges on the materials' reliabilities, and are to date commercially relevant. However, below 100 °C, the performance and life spans of the LTWEs and LTFCs not only depends on the intrinsic kinetics of reactions, but are also highly influenced by the mass transfer in these systems. The latter presents one of the major challenges hindering operations at higher current densities. As shown in **Fig. 2** (b) – (d), the produced gas bubbles and water droplets during operations may be accumulated within the porous electrodes and are not removed from the flow channels.[27,37–39] Such inadequate bubbles and droplets management not only blocks electrochemical reactants lowering catalyst utilization, but also increase local mechanical and thermal stresses accelerating components' degradations [see **Fig. 2** (e) – (g)].[40–43] Understanding and managements of fluid transport are therefore critical for improving designs and operation schemes of LTWEs and LTFCs [see **Fig. 2** (h) – (i)] [29,30,44], overcoming the fluid transport limitations, thereby achieving reliable and efficient LT hydrogen energy systems at industrial-scales.

For this purpose, analytical technologies are greatly demanded for enriching quantitative insights into the relations of LTWEs and LTFCs performance and the multi-scale fluid transport across their components. Conventional optical and radiographic flow measurement methods have greatly advanced understandings on bubble and water dynamics.[45,46] Combined with appropriate electrochemical characterizations methods e.g., electrochemical impedance spectroscopy (EIS) [47,48], these approaches clarified how fluid transport affects performance in small-scale setups, providing design guidelines towards desirable transport behaviors.[49,50] Aiming at industrially-relevant LTWE and LTFC systems, recent researches have evolved towards experiments in larger-scale cells addressing technically relevant components under realistic operating conditions, which is however hard to be investigated with conventional imaging systems. This gap motivated the developments of novel analytical systems, exploring mainly two alternative strategies. On one hand, integrating miniaturized sensors with real LTWE and LTFC stacks, fluid transport related quantities can be directly measured, such as the electrical current, humidity, flow rate, and chemical stoichiometry.[51] On the other hand, alternative physical effects, such as the ultrasonic waves and magnetic fields [52], show great potential to resolve the bubble and water dynamics in practical LTWEs and LTFCs.

Despite these advances, no review has systematically summarized analytical methods towards quantifying fluid transport at the device and system levels. This work aims to address this gap by establishing a unified framework for investigating fluid transport in LTWEs and LTFCs, bridging measurement technologies with the electrochemical performance. We begin with an outline of fluid transport induced limitations in state-of-the-art LTWEs and LTFCs, thereby clarifying the research questions from an energy perspective. Furthermore, the capabilities and limitations of conventional imaging tools for investigating fluid transport in small prototypes are thoroughly evaluated, providing crucial benchmarks defining the requirements for measurement techniques targeting on large-scale systems. Based on this foundation, we highlight recent advances in miniaturized sensors and emerging ultrasonic approaches that open new horizons towards investigating fluid transport under technologically relevant conditions. Finally, we summarize the applicability of existing analytical techniques for specific transport phenomena, and formulate future multi-disciplinary research directions where innovative analytical tools could faithfully contribute to transfer hydrogen energy technologies from laboratory prototypes to reliable industrial-scale deployments.

## 2. Fluid transport in LTWEs and LTFCs

This section provides an overview of the specific fluid transport phenomena in state-of-the-art LTWEs and LTFCs that directly affect their performance. The goal is to provide the necessary background assisting energy researchers to identify the most critical transport phenomena, while guiding instrumentation engineers in designing measurement principles tailored to these electrochemical challenges.

### 2.1. Overview of modern LTWEs and LTFCs

The working principles and the electrochemical reactions of LTWEs and LTFCs addressed in this paper are schematically given in **Fig. 3**. The PEM devices use perfluoro sulfonic acid (PFSA) based membrane e.g., Nafion® [53] as solid-state electrolyte.[54] In PEM water electrolyzers (PEMWEs), pure water supplied to the anode (and optionally to the cathode) is split into $H_2$ and $O_2$ via the hydrogen evolution reaction (HER) and oxygen evolution reaction (OER), respectively [see **Fig. 3** (a)]. This process is reversed in PEM fuel cells (PEMFCs), in which electricity and water are generated by the spontaneous hydrogen oxidation reaction (HOR) and oxygen reduction reaction (ORR) [see **Fig. 3** (b)]. These reactions in PEM devices rely on precious Platinum group metal (PGM) catalysts. In contrast, the alkaline devices, namely AWEs, AEM water electrolyzers (AEMWEs) and AEM fuel cells (AEMFCs) use cost-effective transition metal catalysts.[36] **Fig. 3** (c) & (d) give the corresponding reactions in alkaline environment.[36]

Moreover, the difference of AWEs and AEMWEs lies mainly on the electrolyte. AWEs use high concentration alkaline solution electrolyte together with a non-conducting porous diaphragm e.g., polyphenylene sulfide PPS to conduct ions and prevent from short circuit and gas crossover. In contrast, AEMWEs employ a solid-state electrolyte with high hydroxide ($OH^-$) conductivity, which allows operation with only dilute alkaline feed, reducing system cost and complexity.[36] Similarly, AEMFCs operate with $OH^-$ as the charge carrier, enabling the use of non-PGM catalysts.

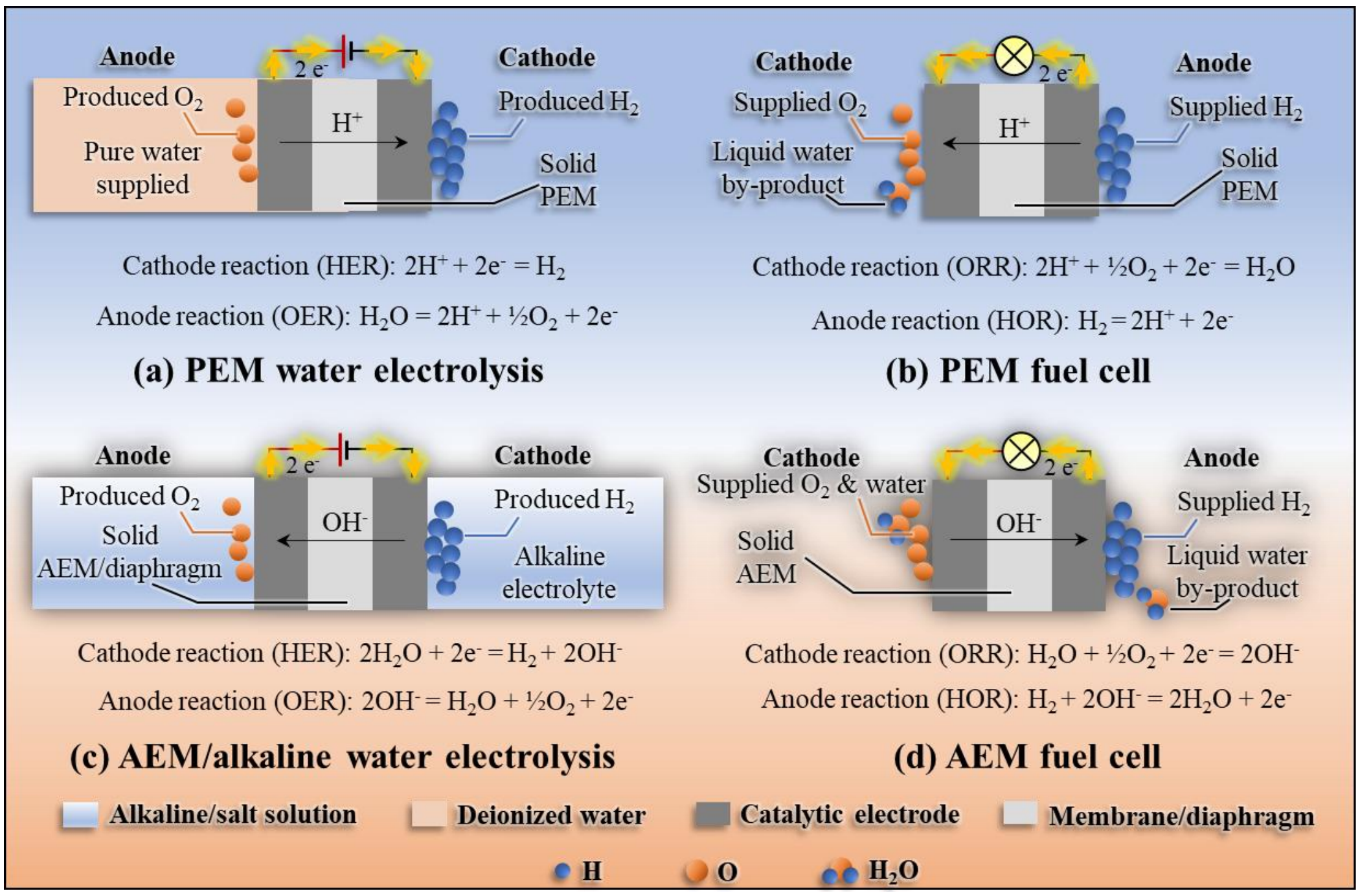

**Fig. 3. The working principles of modern LTWEs and LTFCs:** (a) PEMWE, in which the water supply to the cathode side is optional [55] (b) PEMFC, (c) AWE/AEMWE, and (d) AEMFC.

Commercial LTWE and LTFC stacks usually consist of multiple serially connected flow cells, each with a heterogeneous zero-gap architecture and an active area over 100 $cm^2$, as shown in **Fig. 4** (a). Within a single flow cell, the bipolar plates (BPPs) provide electrical conduction, mechanical support, and flow channels for reactant delivery and (by-)product removal. The typical widths of the channels and ribs, as well as the depths of the channels are in the range of 0.5 mm - 2 mm.[56–58] For enhanced mechanical strengths and gas sealing, metallic BPPs made of Titanium (Ti) and stainless steel are usually employed in commercial LTWEs.[59] In addition to the metallic materials, graphitic composites based BPPs are also often used in LTFCs.[60] The thicknesses of the metallic and composite BPPs [see **Fig. 4** (b)] are usually in the ranges of 75 μm - 500 μm and 1 - 3 mm, respectively.[61–63] Several different flow field designs, such as parallel straight, serpentine, and interdigitated channels [see **Fig. 4** (c)], are usually addressed in commercial LTWEs and LTFCs, directly influencing the fluid transport behaviours.[56]

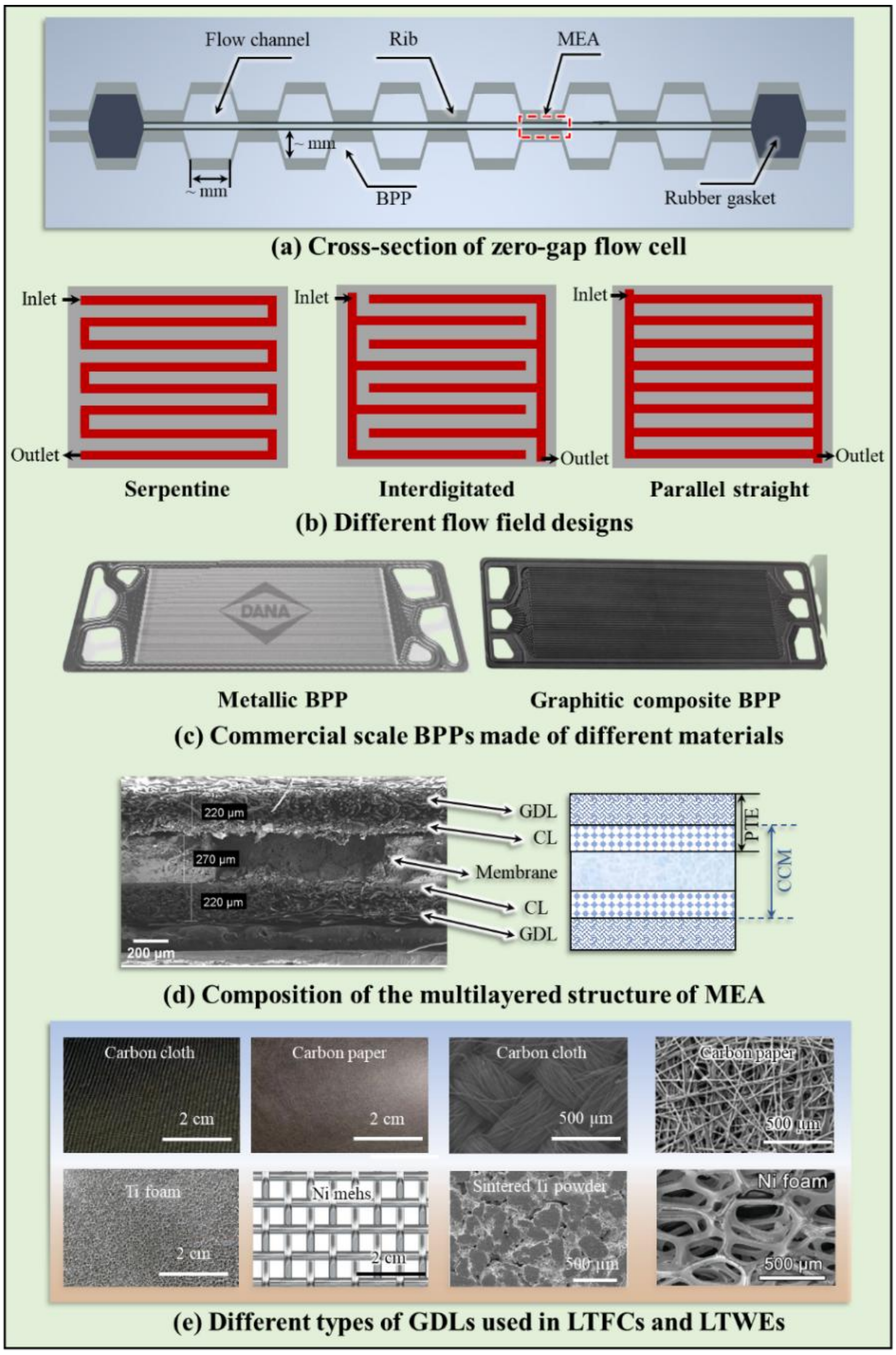

**Fig. 4. The structures, key components and materials of state-of-the-art zero-gap LTWEs and LTFCs.** (a) A cross-sectional view of zero-gap flow cell (not to scale), in which the geometries of the cell components are also given. (b) Different flow field designs employed in LTWEs and LTFCs. [64,65] (c) Commercial scale BPPs made of different materials. Images are adapted from [66,67]. (d) Schematic diagram and SEM image demonstrating the multilayered MEA structure. Image adapted from [68]. (e) The Optical and SEM images of the different types of GDLs in LTWEs and LTFCs. Images are adapted from [69–71].

The membrane electrode assembly (MEA) sandwiched between the two BPPs is the reactive heart of the devices. A MEA typically consists of a membrane, and a porous catalyst layer (CL) with a gas diffusion layer (GDL) at both the cathode and anode sides. The CL enable the redox reactions, and the GDL is used to distribute and remove the reactants and products. Moreover, CLs can be fabricated either by depositing catalyst directly on the membrane or on the GDL [see **Fig. 4** (d)]. The two configurations are termed as catalyst-coated membrane (CCM) [72–75] and porous transport electrode (PTE) [76], respectively. The GDL is another important component, which is also known as porous transport layer (PTL) or liquid/gas diffusion layer (LGDL) in LTWEs. **Fig. 4** (e) shows the commercially available GDLs and the microstructures of these porous mediums. In LTFCs, the GDLs are commonly woven cloths or papers based on Polyacrylonitrile (PAN) carbon fibers, optimized for gas transport and electrical conductivity.[77–79] In LTWEs, PTLs are manufactured from corrosion-resistant metals such as Ti (for PEMWEs) [80] or Ni (for AEMWEs/AWEs) [69,72], and are produced as sintered powders [81], felts [69,72], meshes [62], and open-cell foams [73]. For clarity, in this review we use the terms PTL, GDL, LGDL, and PTE interchangeably when discussing porous mediums, while emphasizing their functions depending on the specific LTWE or LTFC system.

## 2.2 Fluid transport in LTWEs and LTFCs

This section outlines the key fluid transport phenomena in LTWEs and LTFCs, as well as their impacts on system performance.

### 2.2.1 Fundamentals principle of two-phase transport

The evolution of electrolysis generated bubbles undergo three stages, namely the nucleation, growth and detachment [49], as illustrated in **Fig. 5** (a). Bubble nucleation occurs once the local gas concentration ($Cg$) exceeds the saturation concentration of the gas in the liquid ($Csat$), usually at certain positions near the reaction sites where catalysts present.[82] Bubbles grow by absorbing dissolved gas [83], until upward forces e.g., buoyancy [84], flow shear [85], and Marangoni convections [86–88] overcome the downward surface tension, leading to detachment.

In parallel, water exists as vapor, liquid and solid i.e., ice phases in LTWEs and LTFCs.[89] However, in the context of fluid transport, only vapor and liquid phases are typically considered, which are governed by condensation and evaporation processes.[90] As in any mass transfer process, the water transport is primarily driven by diffusion and convections.[91] In general, the former is driven by concentration gradients following the Fick's law, and the latter is induced by pressure gradients. **Fig. 5** (b) depicts the water transport across a MEA: the diffusion coefficient of water molecules ($D_\lambda$) in a MEA depends on the membrane's hydration level ($\lambda$) and the tortuosity and porosity of the GDLs [92,93]; the convection may be driven by various hydrodynamic forces across the MEA ($\Delta p$) [see blue arrow in **Fig. 5** (b)]; in ion-conducting membranes i.e., PEM and AEM, electro-osmotic drag (EOD) further couples water flux to ion migration.[90] The EOD is the phenomenon of water molecules being carried through the membrane along with ions.[91] As illustrated by the black arrow in **Fig. 5** (b), the flux of water transport due to EOD ($j_{EOD}$) depends on the applied current density ($j$), the Faraday's number ($F$), and the EOD coefficient ($\xi_\lambda$) that means the number of water molecules being carried together with each ion through a membrane [91], which is highly sensitive to both the temperature and membrane hydration.[94]

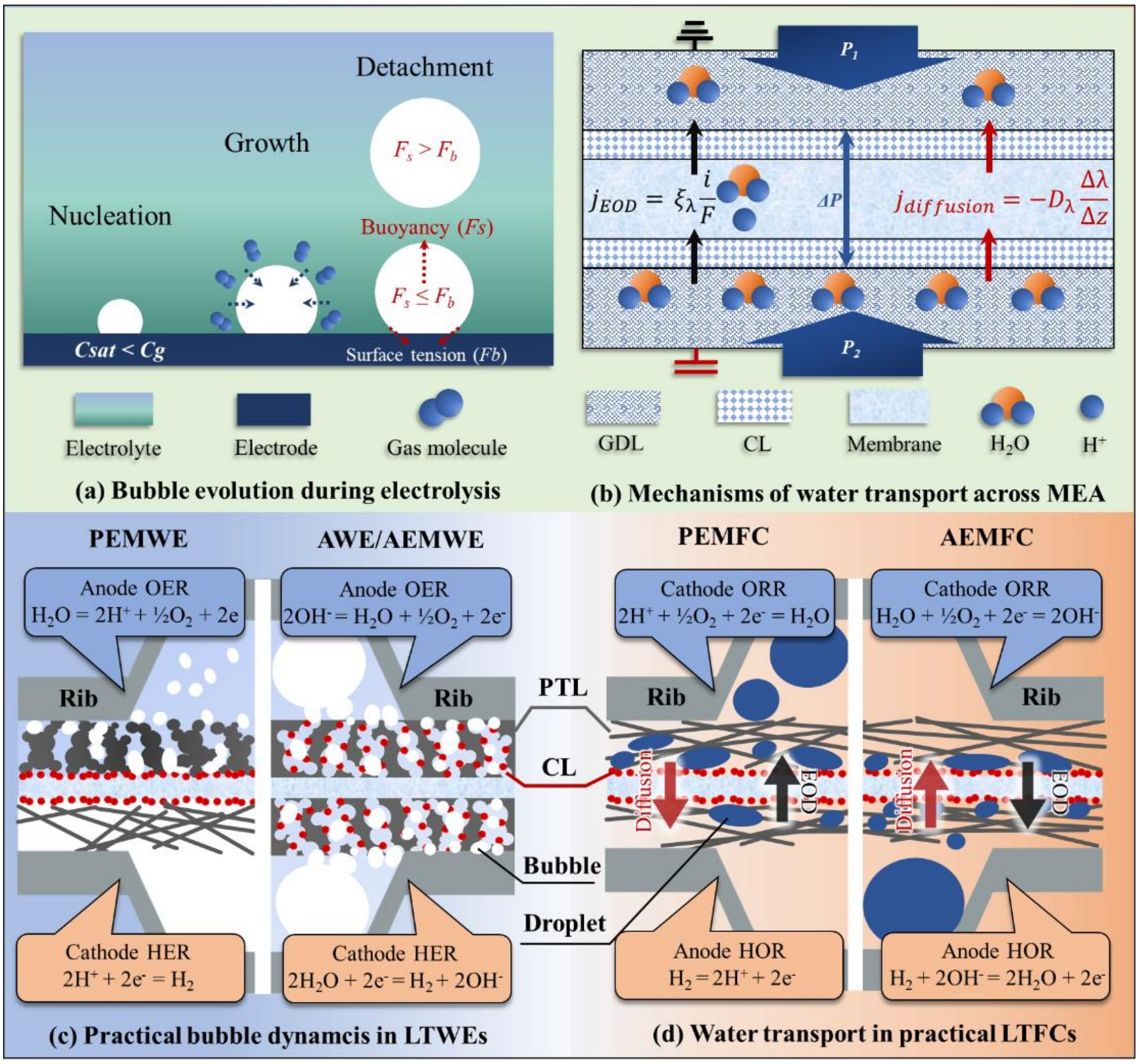

**Fig. 5. Bubble and water dynamics in LTWEs and LTFCs.** (a) Schematic diagram of bubble evolution undergoing the nucleation, growth and detachment. (b) The mechanisms of water transport across a MEA, namely the EOD, pressure forced hydraulic penetration and diffusion, in which *j* stands for the flux of water transport. (c) Bubble dynamics in typical LTWEs. In this illustration, the PEMWE (dry cathode operation) is equipped with sintered powder and carbon fiber type PTLs at the anode and cathode sides, respectively. While, the open cell foam type self-supported PTEs are illustrated for the AWE/AEMWE, in which the bubble evolution and transport are hybrid. (d) Water dynamics in typical LTFCs.

### 2.2.2 Fluid transport in LTWEs

**Fig. 5** (c) demonstrates the bubble dynamics in practical LTWEs. In a PEMWE, the $O_2$ bubbles nucleate at the anode CL due to the OER, while the $H_2$ bubbles may be absent at the cathode under dry cathode operations. In an AWE or an AEMWE, the $H_2$ and $O_2$ bubbles nucleate inside the PTEs at the cathode and anode sides, respectively. The bubbles continue to grow and transport inside the PTEs until they reach the PTE/flow channel interfaces. It is worth to mention that the bubble dynamics in some self-supported PTEs that are entirely coated with catalysts show a hybrid evolution and transport fashion, as shown in the righthand side illustration of AWE/AEMWE in **Fig. 5** (c).[49,72] Within a PTE, the bubble growth and transport are constrained by the pore network, and forced by the complicated conventions. After reaching the PTE/flow channel interface, the bubbles can be released. Experimental investigations suggest that both the bubble transport pathways in the PTEs and the release process are influenced by the operating parameter of a LTWE e.g., the current density [95,96], and configurations of a PTE e.g., the contact angle and pore size [97]. Moreover, the gas bubbles may be accumulated under the ribs, due to the absences of lateral transport pathways, leading to performance losses.

### 2.2.3 Fluid transport in LTFCs

State-of-the-art LTFCs are typically fed with humidified gas for two reasons. Firstly, water vapor hydrates the PEM or AEM membrane materials, enhancing their ionic conductivity and reducing ohmic resistance.[37,98] Secondly, water is directly involved in the ORR at the AEMFC cathode. Furthermore, additional water is produced at the cathode in PEMFCs and at the anode in AEMFCs. The possible water transport pathways in the two types of LTFCs are briefly summarized in **Fig. 5** (d). In a PEMFC, the produced water may diffuse from the cathode CL to the anode side. Meanwhile, due to the EOD, water can also be transported from the anode to the cathode. In contrast, in an AEMFC, EOD causes water to move from the cathode to the anode, but water may also diffuse back to the cathode due to higher water activity on the anode side. As a matter of fact, the exact water contents and distribution in an LTFC is greatly influenced by its operating parameters e.g., mass flow rate and temperature [90], as well as its design configurations e.g., the flow field pattern.[30,42,89] Despite these complexities, the persistent presence of liquid water during LTFC operation has been experimentally verified.[99] Two frequently addressed inappropriate water transport in LTFCs are anode drying and cathode flooding in PEMFCs, as well as cathode drying and anode flooding in AEMFCs. These conditions reduce ionic conductivity, block reactant gas accessing catalysts, and cause strong output power fluctuations, especially at high current densities.[100,101]

### 2.2.4 Impact of fluid transport on cell performance

Inadequate fluid transport management can substantially limit the energy conversion efficiencies of LTWEs and LTFCs, mainly due to increased overpotentials required to maintain operations at high current densities. The excess voltage, beyond the thermodynamic voltage, is known as overpotential. The theoretical cell voltage for water electrolysis and fuel cell operations under standard conditions i.e., 25 °C and 1 atm is 1.23 V. In electrolyzers, this is the minimum voltage required to initiate water splitting, while in fuel cells, it is the maximum voltage obtainable from the electrochemical recombination of $H_2$ and $O_2$. However, overpotentials in practical operations lead to the necessity of higher applied voltages for LTWEs and lower output voltages (for LTFCs), thus reducing the overall energy conversion efficiencies.[102]

Since the Butler–Volmer equation presents the relation between overpotential and current density in redox reactions, it serves as a framework for analyzing fluid transport induced overpotentials in LTWEs and LTFCs. Eq. (1) presents a simplified form of the Butler–Volmer equation that is valid at high overpotentials, and thus suitable for interpreting high current density conditions.[103]

$$\frac{i}{ECSA} = j_{ECSA} = j_0 \cdot \left[\frac{c_{surface}}{c_f} \cdot exp\left(\frac{\alpha \cdot z \cdot F \cdot \eta}{RT}\right)\right] \quad (1)$$

In (1), $i$ is the total current, $j_{ECSA}$ is the current density normalized by the electrochemical active surface area (ECSA) of a catalytic electrode, $\eta$ is the total overpotential, and $c_{surface}$ and $c_f$ stand for the species concentrations at the CL surface and in the flow field far from the CL surface [see **Fig. 6** (a)], respectively. Moreover, $j_0$ is the exchange current density, $z$ presents the number of electrons involved in the electrode reaction, $F$ is the Faraday constant, $T$ is temperature, $R$ is the universal gas constant, and $\alpha$ is the charge transfer coefficient.

Inadequate fluid transport induces three types of overpotentials, namely ESCA reduction induced overpotential, concentration overpotential and Ohmic loss. Firstly, during electrolysis, a part of the reaction sites on the CLs of a LTWE may be electrochemically deactivated, due to coverages of the nucleated bubbles, as shown in **Fig. 6** (a). Likewise, the ECSA of an operating LTFC may also be reduced by the condensation of liquid water on the CLs. As a result, when LTWEs and LTFCs are operated at constant current values, the decreased ECSA due to electrode coverage ($\Theta$) forces to increase local current densities at the rest of the active reaction sites. Therefore, according to (1), an additional overpotential ($\eta_{coverage}$) is necessary to maintain such constant current operations. This is more clearly formulated by the Tafel equation (2), which is derived from the simplified Butler–Volmer equation, showing the logarithmical rise of overpotential with increasing local current density.

$$\begin{cases} \eta = \dfrac{2.303 \cdot R \cdot T}{\alpha \cdot z \cdot F} \cdot log_{10}\left(\dfrac{1}{1-\Theta} \cdot \dfrac{j_{ECAS}}{j_0}\right) \\ \eta_{coverage} = \dfrac{2.303 \cdot R \cdot T}{\alpha \cdot z \cdot F} \cdot log_{10}\left(\dfrac{1}{1-\Theta}\right) \end{cases} \tag{2}$$

Eq. (2) shows that the overpotential $\eta_{coverage}$ rises logarithmically with the increasing of local current density due to the reduced active area, and the Tafel slope i.e., $\frac{2.303 \cdot R \cdot T}{\alpha \cdot z \cdot F}$ corresponds to the value of $\eta_{coverage}$ at a coverage ($\Theta$) of 0.9. For a clear perception, **Table I** lists the values of Tafel slopes for common catalyst electrodes in LTWEs and LTFCs, assuming a charge transfer coefficient ($\alpha$) of 0.5.

**Table I. Typical Tafel slopes of common reactions in LTWEs and LTFCs [a]**

| Reactions | Catalysts | Devices | Z | Tafel slopes ($\eta_{coverage}$ @ $\Theta = 0.9$) |
|---|---|---|---|---|
| **HER** | Pt | PEMWE | 2 | ~ 30 mV/dec |
| | Ni-based | AWE/AEMWE | 2 | 100 – 120 mV/dec |
| **OER** | $IrO_2/RuO_2$ | PEMWE | 4 | 40 – 60 mV/dec |
| | Ni-based | AWE/AEMWE | 4 | 50 – 70 mV/dec |
| **ORR** | Pt | PEMFC | 4 | 60 – 70 mV/dec |
| | Ni-based | Alkaline | 4 | ~ 120 mV/dec |
| **HOR** | Pt | Acidic | 2 | 30 – 40 mV/dec |
| | Ni-based | Alkaline | 2 | 80 – 120 mV/dec |

[a] T = 298.15 K (25 °C), R=8.314 J/mol·K, F=96485 C/mol

**Table I** means that, if 90% of the ECSA is deactivated i.e., $\Theta = 0.9$, the current density at active sites increases by a factor of 10, resulting in an extra overpotential of 30 – 120 mV per decade of current density. If the active sites are fully covered i.e., $\Theta = 1$, no reaction can occur, regardless of applied voltage. This illustrates the significant impact of electrode coverage on overpotential.

Secondly, the presences of bubbles in LTWEs and liquid water in LTFCs forces active species to follow more tortuous diffusion paths through the GDL, as depicted by the dash lines in **Fig. 6** (a). At high current densities, the reactants are quickly consumed at the CLs, and cannot be timely replenished by the mass transport from the distal flow field. This leads to a depletion of reactants at the CL surface i.e., the reactants' concentrations at the CLs ($c_{surface}$) is lower than the values in the flow field ($c_f$). According to (1), such a concentration difference builds up an additional voltage across the entire GDL that is termed as the concentration overpotential ($\eta_{concentration}$) or the mass transport overpotential [104,105], as given in (3). The greater the concentration gradient, the higher the overpotential.

$$\eta_{concentration} \propto \frac{RT}{\alpha \cdot z \cdot F} \cdot \ln\left(\frac{c_f}{c_{surface}}\right) \tag{3}$$

Therefore, the highest achievable current (i.e., the limiting current) densities of LTWEs and LTFCs are determined by the rate of transporting reactants to the CLs [106,107], as shown in **Fig. 6** (b).

Last but not the least, an ohmic loss ($\eta_{ohm}$) exists in any electrical circuit. In AWEs and AEMWEs, the produced bubbles displace the highly conducting alkaline electrolyte, which introduces additional ohmic losses ($\eta_{bub_ohm}$) because of the high electrical resistances of the gas bubbles. The $\eta_{bub_ohm}$ simply follows the Ohm's law, as given in (4), in which $V_{bub}$ is the total volume of accumulated bubbles in an operating LTWE.

$$\eta_{bub_ohm} \propto i \cdot V_{bub} \tag{4}$$

However, the $\eta_{bub_ohm}$ may be neglected in PEMWEs, since they are supplied with nonconductive deionized (DI) water. In LTFCs, the ohmic losses are originated from the dehydrated membranes, due to imbalanced water distribution, such as the commonly addressed anodes drying in PEMFCs and cathodes drying in AEMFCs.[100,101]

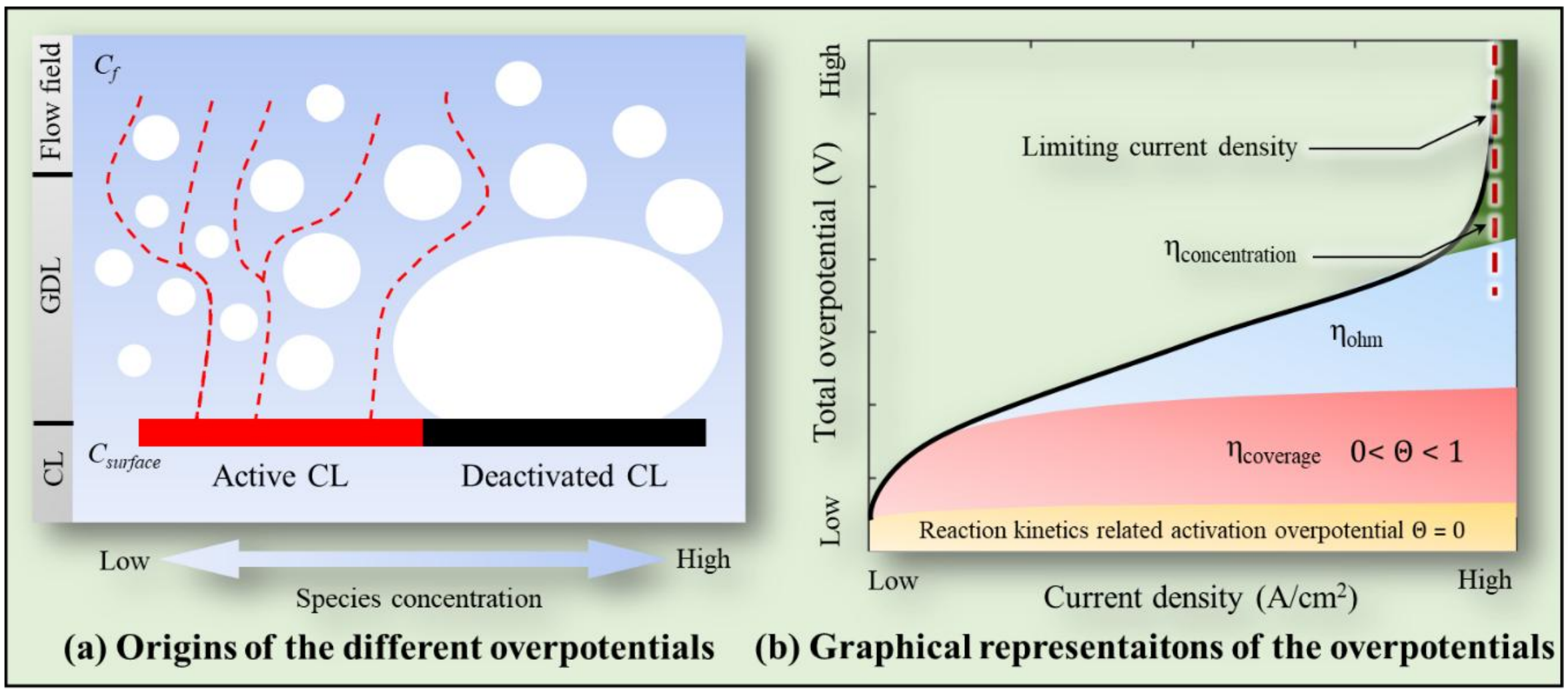

**Fig. 6. The fluid transport induced performance losses of LTWEs and LTFCs.** (a) Schematic representation of the origins of the different overpotentials at an electrode. The white circles indicate the gas bubbles or water droplets in a GDL. The red dash lines stand for the possible transport pathway of reaction species. (b) A graphical representation of the different overpotentials that are derived from the Butler-Volmer equation and the Ohm's law.

In summary, the total voltage loss due to bubble and water transport in a full LTWE and LTFC is given in (5), which is also known as the transport loss.[108]

$$\eta_{transport} = \eta_{coverage} + \eta_{concentration} + \eta_{ohm} \tag{5}$$

Rational engineering of fluid transport is the key to minimize such transport loss, enabling higher efficiencies and expanding the limiting current boundary of LTWEs/LTFCs. Through the analysis in this section, the optimal bubble and water transport behaviours that need to be realized are formulated as follows. In LTWEs, more frequent detachments of bubbles from the PTE at smaller departure diameters are preferred.[49] Furthermore, the bubbles in the flow field should be quickly removed by the supplied water flows. Meanwhile, PTEs should enable greater specific surface area i.e., larger ECSA per unit volume.[49] For optimal performance of LTFCs, water content must be appropriately balanced to avoid water flooding in the flow field, while keeping the membrane adequately hydrated.[50]

### 2.3 Requirements on analytical systems for fluid transport in LTWEs and LTFCs

To realize these optimal fluid transport behaviours necessitates their precise and quantitative measurements, which in turn defines the requirements for analytical systems. More precisely, for LTWEs, the information such as the nucleation positions of bubbles in PTEs, the departure diameters of bubbles from the PTEs, the coverages of PTEs, the volume of bubbles accumulated in the PTEs and flow channels should be spatiotemporally resolved under various operating schemes. Likewise, the spatiotemporal distributions of water content in MEAs and flow fields of operating LTFCs are also of significant importance towards their optimizations.

These measurement scenarios clearly outline the requirements on the analytical systems: they must be able to capture such multiscale fluid transport processes in real-time e.g., >1 Hz, with spatial resolutions ranging from 10’s μm in porous electrodes to 100’s μm in flow fields. In summary, analytical tools that can meet these requirements are the key for connecting fluid transport with practical optimizations of LT hydrogen technologies.

## 3. Optical flow measurement methods for transparent test cells

Conventional analytical techniques for quantifying fluid transport in operating LTWEs and LTFCs primarily rely on imaging methods. Among them, optical imaging systems are arguably the most well-established scientific tools. Assisted with electrochemical characterizations such as cyclic voltammetry (CV) and electrochemical impedance spectroscopy (EIS) [47,48,109–111], these studies contribute significantly for understating the fluid transport induced performance losses, guiding the optimizations of cell design and operating scheme, despite their reliance on transparent accessing windows. This section reviews the major classes of optical methods used for investigating fluid transport in LTWEs and LTFCs, including direct visualization, optical velocimetry techniques, and Schlieren imaging. Their capabilities, limitations and gained insights into LTWE/LTFC optimizations are also discussed.

## 3.1 Direct optical imaging

### 3.1.1 Optical imaging of fluid transport in the flow field

For understanding the overpotentials due to catalyst coverage, direct optical imaging presents an intuitive and straightforward approach for investigating the presences and movements of bubbles and droplets *on the surfaces of GDLs and in the flow fields of transparently designed LTWE and LTFC test models*. Using high speed cameras equipped with microscopic optics, optical imaging can be realized in millisecond level real-time, providing high spatial resolutions down to the diffraction limit of optical waves i.e., sub-micrometer down to wavelength level. A typical setup is shown in **Fig. 7** (a).

In LTWE related studies, the relations of bubble dynamics e.g., bubble departure sizes, the electrode coverages, and the electrodes' configurations i.e., structure, porosity and wettability have been investigated.[108,112–114] These studies conclude that hydrophilic i.e., aerophobic electrodes are able to decrease the bubbles' departure sizes due to smaller surface tension forces [see **Fig. 7** (b)].[113,114] Moreover, electrodes that can enable bubble departure diameters smaller than their pore size are beneficial for minimizing the transport overpotential [108], as shown in **Fig. 7** (b). While, optical imaging in LTFCs related studies revealed that hydrophobic GDLs and higher gas flow rates can facilitate the detachments of water droplets into the flow channels, resulting in better water removal performance.[99,115,116]

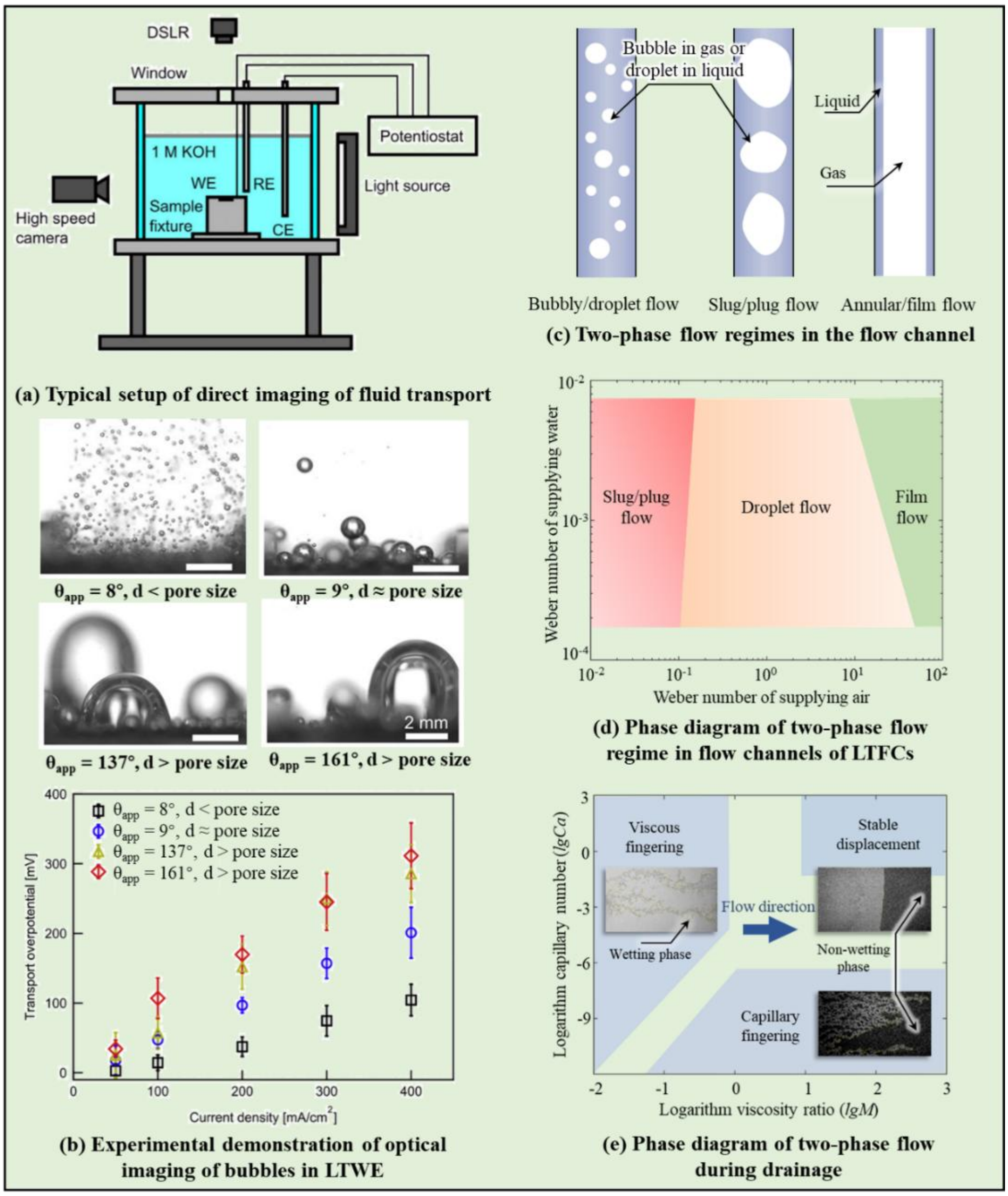

**Fig. 7. Direct optical imaging of fluid transport.** (a) Typical setup of optical imaging of fluid transport in transparent test flow cells. Image adapted from [108]. (b) Experimental demonstration of the relations among the electrode's apparent contact angle ($\theta_{app}$), detached bubble size and the transport overpotential in LTWE. Image adapted from [108]. (c) Schematic illustration of three gas/liquid two-phase flow regimes in the flow channels of LTWEs and LTFCs. (d) The two-phase flow regimes with respect to gas/liquid velocities in flow channels of LTFCs. Data from [117]. (e) The phase diagram of drainage representing gas transporting in LTWEs and water transport in LTFCs, in which the flow regimes are characterized by the displacing front. The phase boundaries are based on [118]. Image of the flow regimes are adapted from [119].

In practice, simultaneous gas and liquid transport in LTWEs and LTFCs results in complex immiscible two-phase flows within both the GDLs and the flow fields.[46,99] **Fig. 7** (b) depicts the three typical two-phase flow regimes that are commonly observed in the channels of LTWEs and LTFCs, namely the bubbly/droplet flow, slug/plug flow and annular/film flow. With optical visualizations, the transitions between these different flows regimes with the changes in the flow rates and the operating current densities were investigated. In general, increasing the reactant flow rate helps to reduce the bubble/droplet size in the flow channel and supress the transitions from bubbly/droplet flows to slug flows [25,99,120], thereby mitigating the water flooding in LTFCs and the bubble accumulations in LTWEs.[121] In addition, *ex-situ* experiments with precise flow rate controls were conducted to determine the superficial velocity boundaries of gas and liquid that result in the different flow regimes. A general phase diagram illustrating these flow regimes in LTFC flow channels is shown in **Fig. 7** (c) [122,123], in which the superficial velocities are expressed by the dimensionless Weber numbers [a]. This diagram presents a unified relation between flow regime and operating parameters, providing valuable insights into optimizations of flow field designs and control strategies that minimize water flooding and gas trapping, thereby improving the performance of LTWEs and LTFCs.

### 3.1.2 Optical imaging of fluid transport in the porous electrodes

It is also crucial to understand and optimize the fluid transport in the GDLs of LTWEs and LTFCs, which is however challenging for direct optimal imaging, due to the limited penetration depth of visible light through such opaque materials.[124] To overcome this limitation, optically transparent 2D or 3D porous models have been designed and fabricated, commonly via techniques e.g., photolithography, to mimic the morphology and properties of actual GDLs.[124–127] These artificial models enable detailed investigations of fluid transport in porous media with different flow velocities, fluid viscosities, surface tensions, and contact angles.[128–130]

In LTWEs and LTFCs, fluid transport is typically represented by a so-called "drainage" process, where a non-wetting phase (i.e., gas in LTWEs and water in LTFCs) displaces a wetting phase (i.e., water in LTWEs and gas in LTFCs) within a porous network. Using optical imaging in the transparent models, researchers have mapped the resulting flow regimes into a generalized phase diagram. In **Fig. 7 (e)**, the drainage phase diagram is expressed with respect to the dimensionless capillary number (Ca) [b] and the viscosity ratio (M) [c] between the non-wetting and wetting phases.[118] This phase diagram therefore serves as a valuable tool for elucidating the fluid transport regimes in the GDLs of operating LTWEs and LTFCs. In particular, in the stable displacement regime, the non-wetting phase advances smoothly through the pore network, potentially leading to complete saturation of the GDL by gas in LTWEs, or by water in LTFCs. Such conditions can significantly limit their performance by blocking reactant access. Consequently, this phase diagram also offers practical guidance for engineering the porous structure of GDL materials to maintain favourable fluid transport properties, mitigating water flooding and gas trapping induced performance losses in LTFCs and LTWEs, respectively.

---

[a] $We = \rho \cdot u^2 \cdot d/\sigma$, where $\rho$ is density, $u$ is the superficial velocity, $d$ is the diameter of droplet, $\sigma$ is the surface tension.

[b] $Ca = \mu \cdot u/\sigma$, where $\mu$ is the viscosity of the displacing (non-wetting) phase, $u$ is the superficial velocity, $\sigma$ is the surface tension.

[c] $M = \mu_{non-wetting}/\mu_{wetting}$, where $\mu$ is the viscosity.

### 3.2 Optical velocimetry: resolving convective flows

To study the mechanisms of practical bubble and droplet dynamics, optical flow velocimetry techniques have emerged as powerful tools for investigating convective flows in LTWEs and LTFCs, revealing insights into efficient bubble and droplet removal strategies. These flow velocimetry methods include particle image velocimetry (PIV), particle tracking velocimetry (PTV) and laser Doppler velocimetry (LDV), allowing for visualizing and quantifying local velocity fields by tracking moving fluid structures e.g., bubbles and droplets, or seeded tracer particles from a series of consecutively captured images.

#### 3.2.1 Particle image velocimetry (PIV)

The setup and basic principle of PIV are shown in **Fig. 8** (a). To capture high contrast images of moving particles, a pulsed laser is used to generate an intense illumination sheet at a certain time interval ($\Delta t$). A series of images of the illuminated moving particles are consecutively acquired by a high-speed camera synchronized with the pulsed laser. To determine the displacements of particles between two successive acquisitions, these images are first divided into small so called "interrogation windows", with a size of e.g., 32 pixels × 32 pixels. Cross-correlations of these interrogation areas in the two images are then performed, and the displacement is indicated by the position of the correlation peak [see **Fig. 8** (a)]. Knowing these displacements and the acquisition time interval, the instantaneous velocity vector of each interrogation area can be calculated, generating a dense Eulerian velocity vector map. The spatial resolution of PIV is thus determined by the size of the interrogation window.

With PIV, the velocities of bubbles in narrow vertical flow channels were investigated under different current densities [131], as shown in **Fig. 8** (b) where the spatial resolution was 640 μm. This method was also extended to map the velocity fields of humidified water droplets in LTFCs.[132] However, the motions of bubbles and droplets may not be sufficient to characterize the convective flow fields in LTWEs and LTFCs, as the accuracy of PIV relies on the ability of imaged particles to faithfully follow the fluid motion.[133] For this, neutral buoyant polystyrene particles are often used for investigating water flows in LTWEs.[134] While, for LTFCs, it was proved that the velocity field of water droplets with diameters greater than 1 μm differs significantly from that of the air flow in U-shaped flow channels with Reynolds number [d] higher than 400 [135], as shown in **Fig. 8** (c). This discrepancy may hinder validations of numerical simulations and detailed investigations of complex flow phenomena, such as recirculation and corner flows. To overcome this limitation, smoke seeding particles with diameters smaller than 1 μm were introduced [133], which enabled successful reconstruction of convective flow fields with a Reynolds number of 700, as illustrated in **Fig. 8** (d), where recirculation zones and corner flows are clearly resolved, and a higher spatial resolution of 42 μm was realized in such a smaller area.

---

[d] $Re = \rho \cdot u \cdot D/\mu$, where $\rho$ is the density, $u$ is the superficial velocity, $D$ is the channel diameter, and $\mu$ is the viscosity.

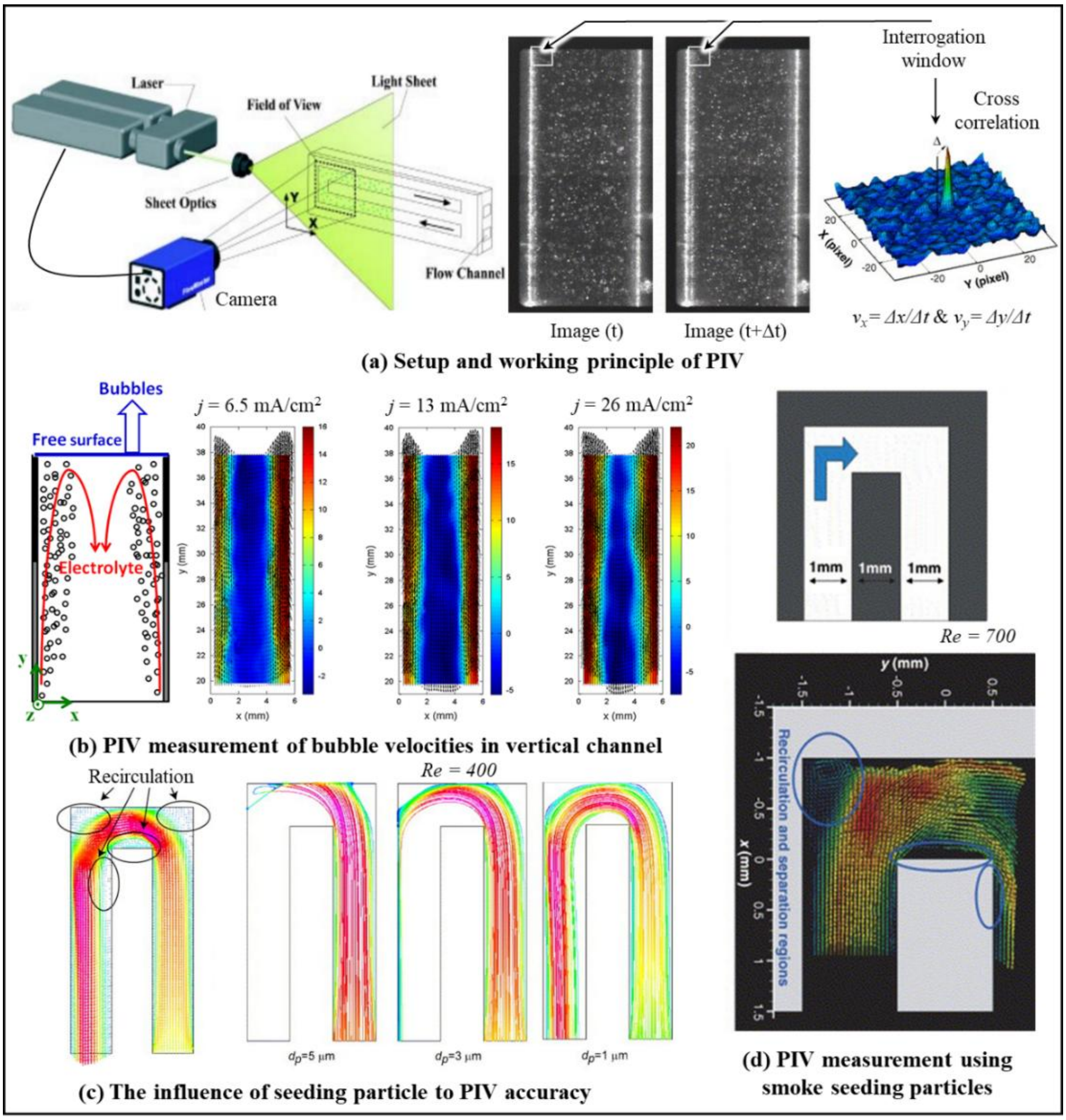

**Fig. 8. Particle image velocimetry (PIV) for LTWE and LTFC studies.** (a) The basic processing in PIV. Image adapted from [132]. (b) PIV investigation of bubble induced convections in vertical flow channels at different current densities. Image adapted from [131]. (c) Droplets larger than 1 μm are not suitable for investigating air flows in LTFCs with Reynolds number (Re) higher than 400. Image adapted from [135]. (d) PIV setup using seeding particles. Image adapted from [133].

### 3.2.2 Particle tracking velocimetry (PTV)

The experimental setup for PTV is largely identical to that used in PIV, as already shown in **Fig. 8(a)**. However, unlike PIV, which calculates average velocities over interrogation areas, PTV tracks the trajectories of individual particles across sequential image acquisitions, thereby offering a Lagrangian view of the flow field. As a result, the spatial resolution of PTV is in theory higher than that of PIV, being limited only by the optical resolution and particle size, rather than by the size of interrogation windows. Furthermore, PTV is also more sensitive to transient and localized flow structures that may be smoothed out or averaged in PIV measurements. These features make PTV particularly advantageous for investigating detailed flow structures near interfaces, such as Marangoni convection around bubbles and droplets in shear flows.[136]

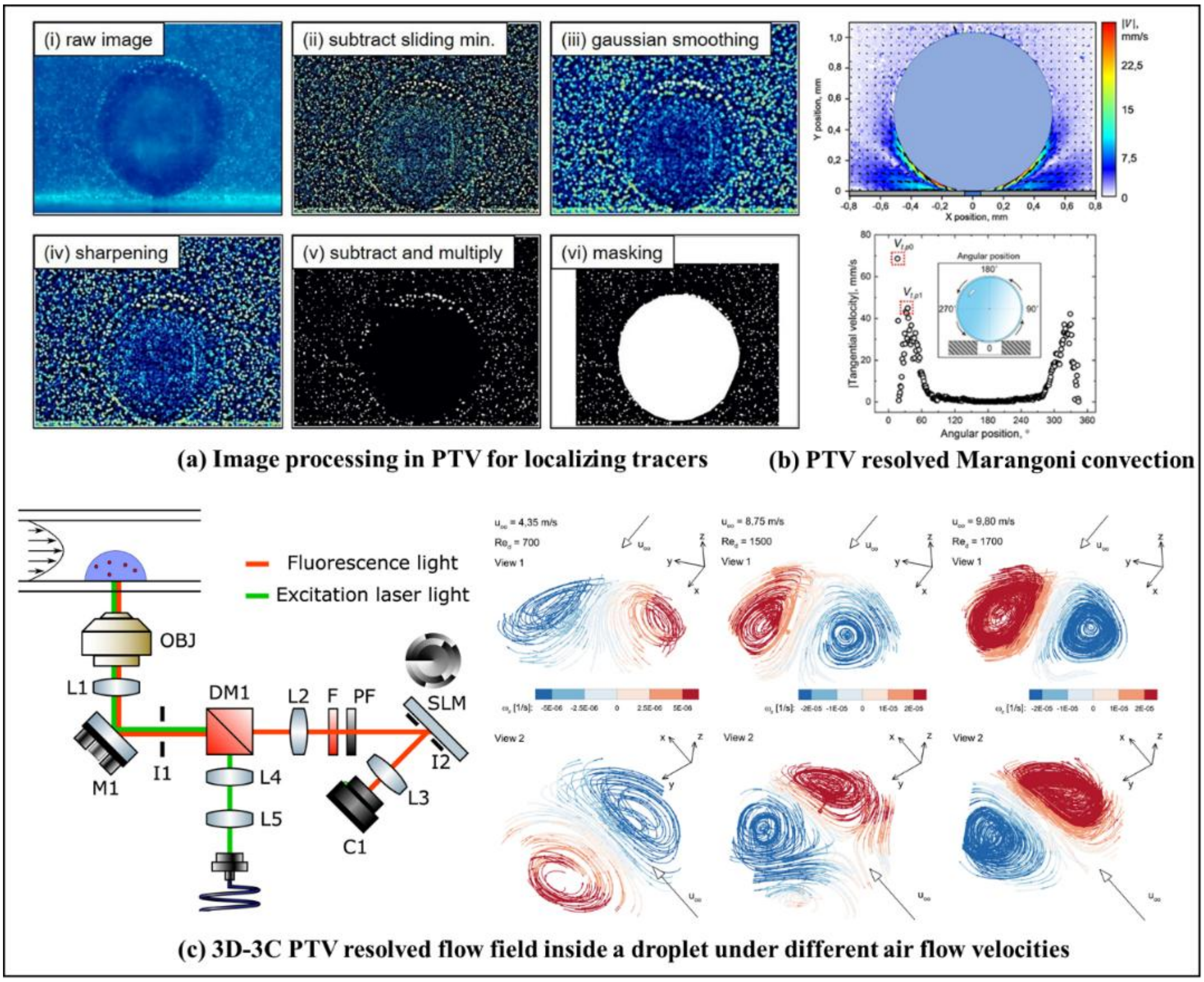

**Fig. 9. Particle tracking velocimetry (PTV) for LTWE and LTFC studies.** (a) Image pre-processing to for localizing centroid of individual tracing particles. Image adapted from [88]. (b) PTV resolved Marangoni convection (red circle) around an evolving bubble on a microelectrode. The upper image shows the velocity field. The lower image shows the tangential velocity around the bubble. Image adapted from [88]. (c) 3D-3C PTV of flow inside water droplet under different air flow velocities in a flow channel. Image adapted from [137].

*A. Babich et al*. used PTV to investigate Marangoni convection around electrolysis-generated $H_2$ bubbles.[88] **Fig. 9** (a) shows the general image processing workflow for localizing the centroid of each tracer particle, enabling faithful tracking with certain algorithms. **Fig. 9** (b) illustrates the velocity field around a $H_2$ bubble evolving on a microelectrode, where convection patterns near the bubble root are clearly observed. In addition to 2D planar measurements, advanced PTV instrumentations have also been developed for 3D-3C (three-dimensional, three-component) flow mapping.[137–140] Using adaptive optics, *C. Bilsing et al.* investigated the flow inside sessile droplets under different air flow velocities [137], as illustrated in **Fig. 9** (c). Furthermore, adaptive optics with aberration correction techniques have been developed to mitigate optical distortions, thereby enabling accurate flow measurements through curved boundaries, such as bubbles and droplets.[139,141]

Other optical velocimetry techniques, such as Laser Doppler Velocimetry (LDV), have also been employed to investigate flow dynamics e.g., around water droplets in LTFCs, achieving high precision with relative velocity uncertainties below 0.1% [142,143], and even enabling simultaneous measurements of flow and temperature fields [144]. These developments significantly enhance the ability to reconstruct complex convective flows in LTWEs and LTFCs, providing insights into practical bubble and droplet dynamics, which can in turn be used for designing efficient bubble and droplet removal strategies. However, *Y. Han et al.* recently found that the use of tracer particles may alter the bubble dynamics.[134] Thus, particle-based measurements for LTWEs must be carefully considered, as the measurement may lead to inaccurate conclusions on device performance.

### 3.3 Schlieren imaging: towards improved understanding on concentration overpotential

To addressing the concentration overpotential, Schlieren imaging, as a tracer-free technique, can directly visualize the variations of spatial changes species concentration during operations of LTWEs and LTFCs, by measuring the refractive index through transparent medium. The German term "schlieren", meaning "streaks", originates from its early use in detecting optical imperfections in glass. The basic setup and working principle of schlieren imaging is illustrated in **Fig. 10** (a). In a typical configuration, a collimated light beam passes through a transparent test cell. When the light encounters regions of varying refractive index, it is deflected from its original path. A knife edge or spatial filter placed at the focal point of a lens or mirror blocks a portion of this deflected light, producing contrast based on the direction and magnitude of the deflection. As a result, otherwise invisible flow structures, such as temperature and concentration gradients, become visible.

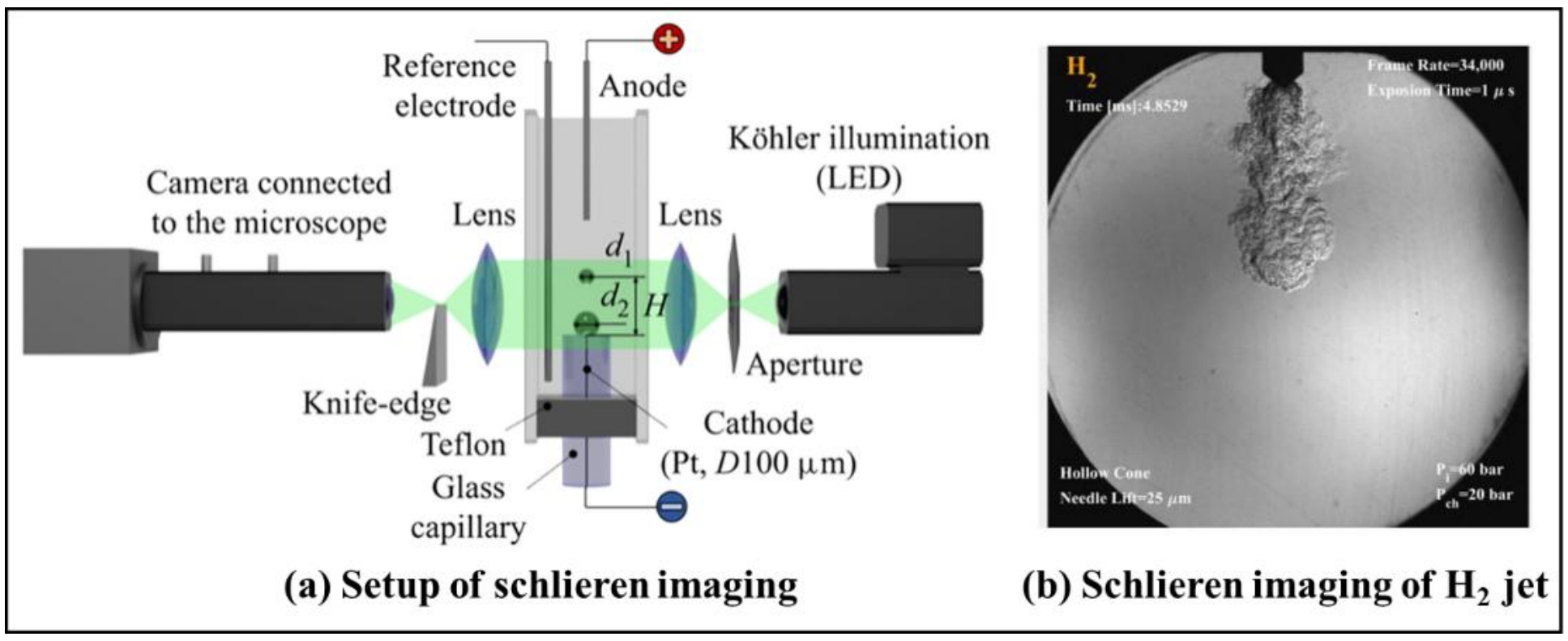

**Fig. 10. Schlieren imaging. (a) The setup and principle of schlieren imaging.** Image adapted from [145] (b) An example of schlieren imaging of $H_2$ jet, analogues to investigations of LTFCs. Image adapted from [146].

With schlieren imaging, the $H_2$ and $O_2$ bubbles dynamics induced temperature and concentration variations have been extensively investigated during water electrolysis.[145,147,148] Although schlieren imaging has not yet been reported in LTFC systems, recent studies on diagnostics of internal combustion engines imply a direct transferability. *M. Yeganeh et al.* reported schlieren visualizations of hydrogen and methane jets [146], providing analogues for understanding gas flow in LTFC channels, as shown in **Fig. 10** (b). *J. Lee et al.* demonstrated the schlieren imaging of fuel vaporization [149], which is equally relevant for visualizing water transport in LTFCs. In short, schlieren imaging is a powerful method providing visualized evidence of bubble and droplet induced mass transport losses, which remain among the least understood phenomena in the operation of LTWEs and LTFCs. However, the temperature and concentration cross sensitivity must be properly addressed.

### 3.4 Summary of optical flow measurement methods

In summary, the optical imaging techniques have yielded fruitful results for providing visualized mechanisms on the influences of fluid transport to the performance of LTWEs and LTFCs. **Table II** outlines these optical measurement methods.

**Table II. Summary of optical flow measurement methods for fluid transport in LTWEs/LTFCs**

| Techniques | Direct imaging | Optical velocimetry | Schlieren imaging |
|---|---|---|---|
| **Primary Output** | Visual recording of visible features | Velocity vectors of fluid flow | Changes in refractive index |
| **Applications** | Bubble/droplet size, position, flow regimes | Convection phenomena, velocities and trajectories of bubbles/droplets | Species concentration and temperature |
| **Key insights** | Flow field geometry, GDL pore network structure, and operating parameters | Convective flow, mechanisms of practical bubble and droplet dynamics | Mechanisms of concentration overpotentials |
| **Complexity** | Low (requires mainly optics, illuminations, and high-speed cameras) | High (requires pulsed laser illuminations, image processing software, precise synchronization, seeding particles) | Moderate (requires collimating/focusing optics, knife edge, precise alignment). |
| **Spatial resolution** | Optical diffraction limit ~ optical wavelength level possible | PIV: interrogation window size limited<br>PTV: optical diffraction and particle size limited | Optical diffraction limit ~ optical wavelength level possible |
| **Other comments** | - Transparent cells and cell components needed | - Careful considerations on seeding particle intrusiveness and laser illumination induced heating<br>- Other methods based on interferometry and laser Doppler also exist | - Careful dealing with vibrations<br>- Temperature and concentration cross sensitivity |

Recent developments on minimally invasive fibre optics further extended the applicability of these methods for investigating fluid transport in flow fields of real operating LTFCs and LTWEs.[150] However, the modifications of the LTWEs and LTFCs with optical accessing windows must be carried out carefully to avoid the influences on the expected fluid transport behaviours.[49] Ultimately, the limited penetration depth of optical waves, even in the terahertz (THz) band [151–154], remains a major hurdle because it hinders the direct *operando* visualization of fluid transport within the opaque GDLs/PTLs, and thereby a complete understanding of fluid transport limitations in practical LTWEs and LTFCs.

## 4. Radiographic imaging for lab-scale prototypes

Comparing with the optical waves, X-rays and neutrons are able to penetrate the opaque BPPs and MEAs, which allows the investigations in test cells consisting of technically relevant materials. These radiographic imaging techniques leverage the physical interactions between X-rays or neutrons and the internal structures of LTWEs and LTFCs to construct spatial distributions of gas bubbles and liquid water. Specifically, the X-rays interact with electrons in the atomic shell, while neutrons interact with atomic nucleus. Moreover, X-rays are EM radiations with high photon energies in a range of 100 eV – 100 keV, corresponding to wavelengths of 0.01 nm – 100 nm.[155] While, neutrons are uncharged subatomic particles, and the equivalent wavelengths of thermal and cold neutrons fall within 0.09 – 0.26 nm and 0.26 – 1 nm, respectively.[156] These short wavelengths enable in principle ultrahigh spatial resolution, enabling insights into the rational designs of the multiscale components in LTWEs and LTFCs. X-rays and neutrons measurements in operating test cells are based on two principles, namely attenuation and elastic scattering effects.

This section highlights both the strengths and the limitations of these radiographic imaging techniques. On one hand, they provide valuable benchmarks and insights under controlled environment in small-scale setups. On the other hand, recognizing their restricted scalability towards commercial-scale systems defines the requirements for developing the innovative analytical tools that are capable of transferring lab-scale insights to industrially relevant implementations.

### 4.1 Radiography and computed tomography (CT)

Based on the attenuation principle, X-rays and neutrons radiography and CT are well developed tools for visualizing gas and liquid distributions, and the setups are shown in **Fig. 11** (a). In radiography, a test cell is located between an incidence beam and a detector, and the incident X-rays or neutrons are attenuated along the propagation path in the test cell, following the Beer-Lambert law.[157,158] Due to the different attenuation coefficients of X-rays and neutrons in the LTWE/LTFC components, water and gas, the intensities of the through transmitted beam are projected as 2D images showing the spatial distributions of bubbles in LTWEs and water in LTFCs.

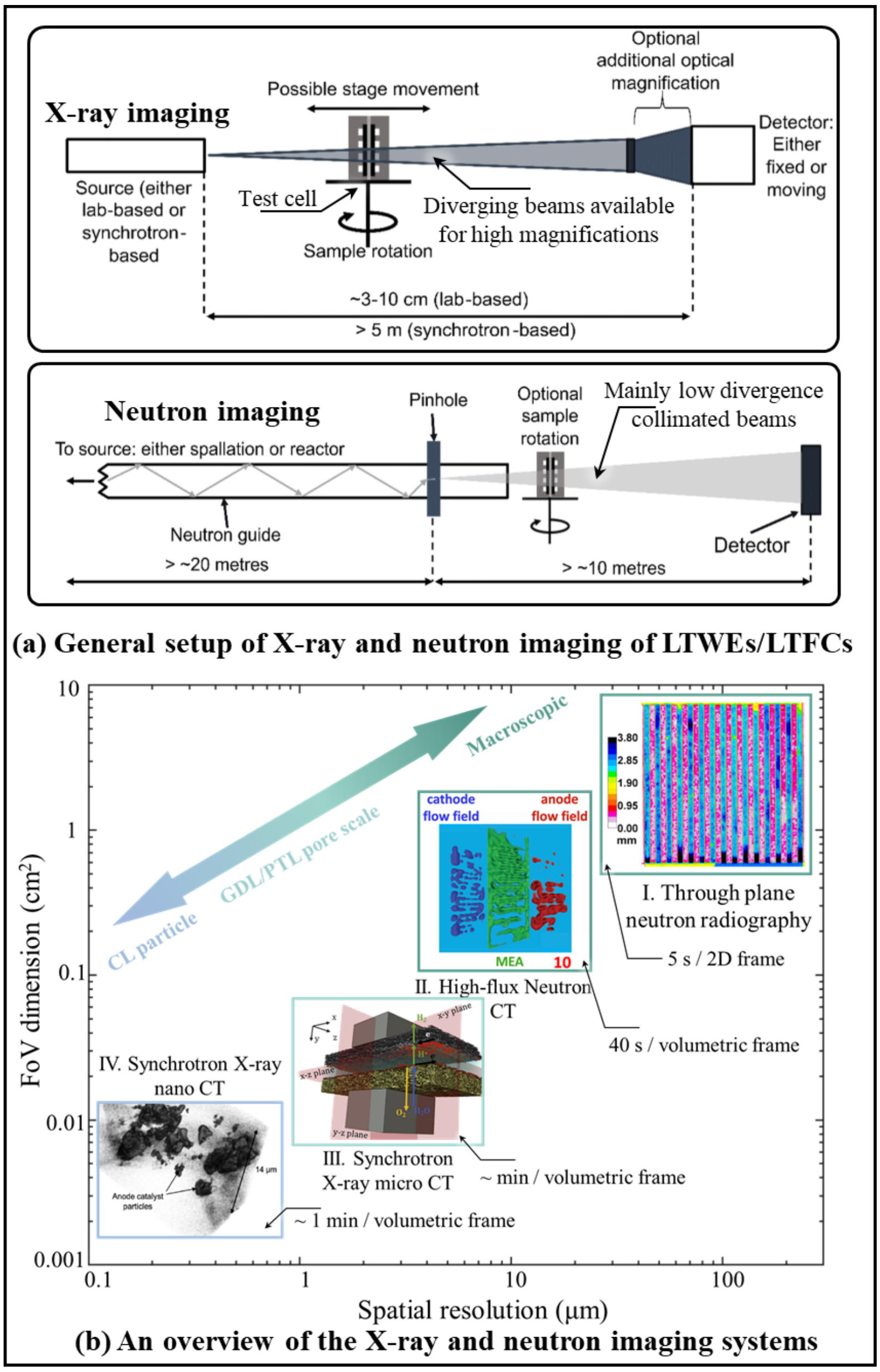

**Fig. 11. X-Ray and neutron imaging fluid transport in LTWEs and LTFCs.** (a) Schematic diagrams of the setups of X-ray and neutron radiography. Image adapted from [159]. (b) An overview of the performance of different X-ray and neutron imaging techniques. Inset I - IV adapted from [95,160–162].

As a 2D imaging technique, the bubble and water saturation levels can be quantified via one-dimensional integral and proper pre-calibrations, either in plane (parallel to the MEA) or through plane (perpendicular to the MEA) [also see inset I in **Fig. 11** (b)].[163,164] To resolve the heterogenous 3D fluid transport, the test cell has to be axially rotated for projecting the incidence beam from different angles, and the acquired signals are then reconstructed using computer aided algorithms e.g., the inverse Radon transform. In general, the performance of an X-ray or neutron imaging system is influenced by multiple factors, including the beam profile and flux, exposure time, thickness and material of a cell under test, and the specifications of the scintillator as an essential component of the image acquisition system.[165] **Fig. 11** (b) summarizes the typical spatiotemporal resolutions and the corresponding FoV dimensions realized by different imaging systems reported in the literatures.[95,160–162]

Due to a lack of focused neutron beams, neutron radiography (NR) is primarily used for investigating gas and water distribution in test cells with large active areas e.g., above 50 cm$^2$.[163,166–169] The NR systems usually provide spatial resolutions at the level of 10's μm to 100's μm, and temporal resolutions at the level of second to minute.[160,169] This is because the low neutron flux results in longer exposure times to achieve adequate signal to noise ratios (SNRs). Taking advantage of modern high-flux beamlines, *R. Ziesche et al.* recently reported the first high speed neutron CT of water evolution in an operating PEMFC test cell with an active area of 2 cm$^2$ where a temporal resolution of 40 s was realized [see the inset II of **Fig. 11** (b)].[161] These visualizations provide significant insights into optimizing the flow field designs [170], selecting the suitable GDLs/PTLs for certain flow field patterns [160,171], and understanding the water and bubble evolution under different operating parameters [172,173]. However, scintillators for neutron detection typically possess large grain sizes for ensuring sufficient SNR, which leads to coarse spatial resolutions of the neutron imaging systems. This hinders the visualization of individual gas bubble and water droplet dynamics, and the investigations of the pore scale fluid transport in the GDLs/PTLs.

In comparison, X-ray imaging can resolve fluid transport at pore scale. Using focused beams, X-ray microscopy and micro CT systems are able to provide (sub)micrometer level high spatial resolutions, albeit within limited FoV of several mm$^2$.[174] The imaging speeds of X-ray systems are determined by the beam characters, in terms of flux and photon energy. Laboratory based low-flux X-ray imaging systems, usually take several seconds to produce a 2D radiography and more than an hour to acquire a 3D tomography of a small test cell.[175,176] While, using high flux synchrotron X-ray beamlines, high resolution 3D imaging at second or sub-second level real-time [see inset III of **Fig. 11** (b)].[96,177] This allows investigating the sizes and frequencies of gas bubbles and water droplets detachments at the GDL/PTL flow channel interfaces [89,117], and their dynamics in the flow channels [178,179]. More importantly, the resolved gas and water transport pathways in the different GDLs/PTLs pave the way for engineering their microstructures towards more efficient mass transport.[95,96,117,180–183] Nevertheless, synchrotron X-ray nano CT systems are able to realize 10's nm level high resolutions for resolving the catalyst particles, as shown in the inset IV of **Fig. 11** (b). However, this extreme spatial resolution comes at the cost of severely restricted FoV of only $10^{-4}$ mm$^2$ – $10^{-3}$ mm$^2$, which significantly compromises the link between pore-scale insights and realistic cell designs.

### 4.2 Small and wide angle scattering techniques

Unlike the morphological imaging of radiography and CT systems, small/wide angle scattering (S/WAS) techniques open new horizons for quantifying gas/liquid saturation at microscales, exploring the elastic scattering of X-rays and neutrons with materials' specific microstructures. In elastic scattering, the changes in the energies of scattered X-rays and neutrons are negligible, while the whole system complies the momentum conservation. As a result, shown in **Fig. 12** (a), the scattered beam is deflected by an angle of 2θ from the incident direction, producing a scattering spectrum characterized by the transferred momentum vector $q$, which encodes the material's unique microstructural information.[184] Such a material microstructure specified interaction therefore allows accurate *operando* quantifications of the membrane hydration [185], CL utilization and degradation due to fluid transport.[186] To enable spatially resolved information, the S/WAS techniques can both be performed in 2D with a point wise scanning fashion [187], and in 3D using CT reconstruction algorithms [188], within larger FoV e.g., above 5 $cm^2$ [185,187,189]. S/WAS measurements uniquely link the microscopic catalyst and membrane states to the macroscopic cell performance, providing practical insights into optimizing cell design and operating condition.

It is however worth to mention that S/WAS instrumentations require specific designs of test cells to prevent the blockage of scattered beams and the parasitic scattering events from other components [118,188]. Moreover, appropriate incident beam energy must be selected to intensify the desired scattering effects.[185,187,189] In addition, the interpretations of acquired spectrums usually require complicated pre-calibration procedures across components at well-defined water saturations.[185].

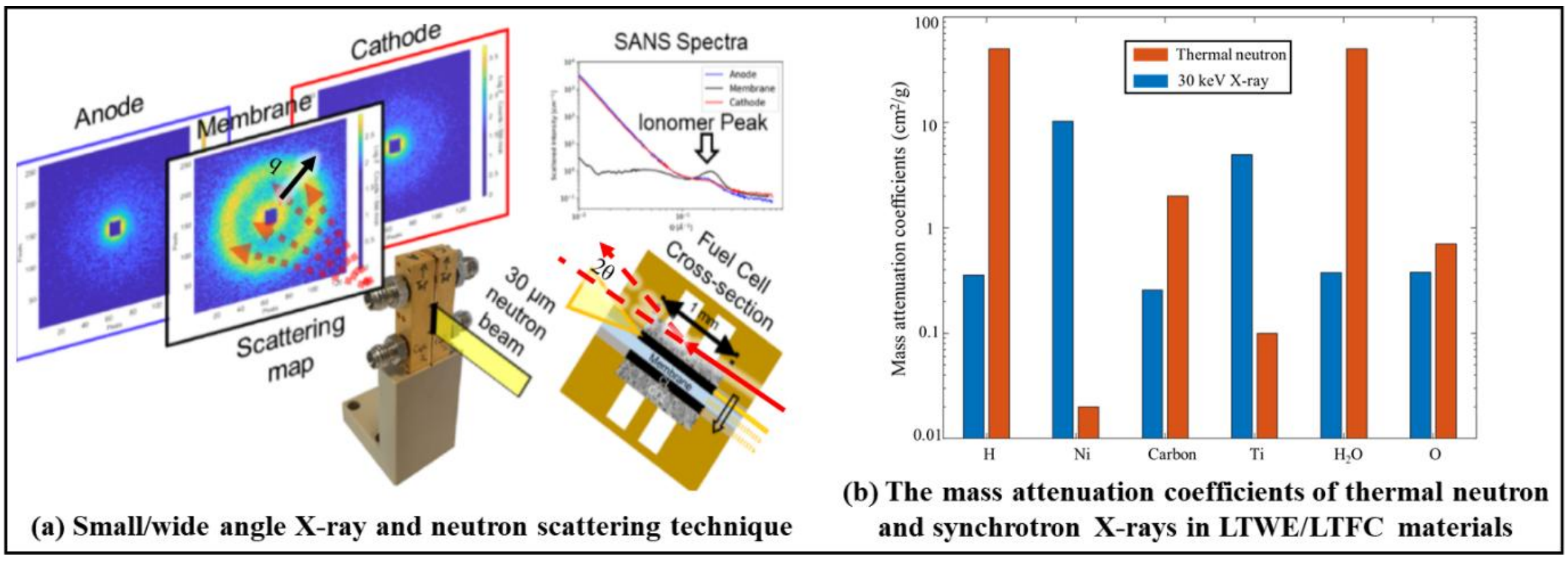

**Fig. 12. X-Ray and neutron imaging systems.** (a) The physical mechanism and experimental setup of neutron and X-ray based small/wide angle scattering techniques. Image adapted from [187]. (b) The mass attenuation coefficients of X-rays and neutrons in the common materials employed in LTWEs and LTFCs. Data adapted from [157,158].

### 4.3 Summary of radiographic methods

Despite the successfully utilities, X-ray and neutron imaging also pose certain limitations. Firstly, in LTWEs related studies, these imaging systems usually show a lack of adequate contrast to resolve the gas/liquid transport in the technically relevant PTLs, e.g., made of Ti and Ni.[95,174] **Fig. 12** (b) plots the linear attenuation coefficients of synchrotron X-rays, with photon energy of 25 keV – 30 keV, and neutrons in the common materials of LTWE and LTFC components, as a measure of the probability of interaction per unit length. In short, X-rays are strongly attenuated in metals [157], resulting in diminished beam intensity and poor gas/liquid contrast in metallic PTLs [96]. To tackle the contrast issue, most studies introduce non-technically-representative carbon based PTLs instead of the metallic ones.[95,119,174,178] On the contrary, neutrons are mostly attenuated by water [158], making metallic PTLs indistinguishable from the evolved gas phase.[160,190] Although the phase contrast issue can be solved by introducing X-ray contrast agents to enhance liquid-phase absorption [96], or using combined X-ray and neutron imaging schemes [96,160], the complex and costly instrumentations, and limited beamline access can significantly hinder lab-based research workflows. Second, the synchrotron X-rays may induce ionizing radiation degradations to the membrane and electrocatalysts, deviating the observed gas/liquid transport phenomena from the realities.[191] In addition, the X-ray and neutron CT systems face significant trade-off between imaging performance and FoV, which is challenging for scaling up test cells.

These limitations collectively pose a major challenge for translating insights obtained from small prototypes to industrial-scale LTWEs and LTFCs. This motivates innovations in analytical systems exploring miniaturized sensors and emerging physical effects towards directly measurements of fluid transport at industrially relevant scales, as discussed in the following sections.

## 5. Embedded micro-sensor networks: probing localized fluid transport events

Recently, sparse sensor array schemes emerge an innovative strategy for investigating LTWEs/LTFCs at the device and system levels. For measuring the internal fluid transport, miniaturized sensors can be integrated with an investigated LTWE/LTFC at different positions of interest, owing to the developments of the micro sensor technologies. The spatiotemporal distributions of these measured physical quantities are then used to obtain relations among cell design, operations and electrochemical performance. To date, such sparse sensor network schemes have achieved long-term (e.g., over 100 h) real-time monitoring of LTWEs and LTFCs at commercial scales, using affordable electronics.[192] This demonstrates that *operando* insights are no longer restricted to small test cells, but can extend into full-size stacks, directly contributing to practical applications. This section reviews the state-of-the-art developments of this scheme, addressing the analysis methods using different types of sensors and sensor integration schemes.

### 5.1 Implanted RH/T sensors: mapping local water content

In order to obtain the water contents in operating LTWEs/LTFCs, the use of humidity sensors provides a direct measurement. Humidity is typically expressed as relative humidity (RH) defined as the ratio of the partial pressure of water vapor and its saturation pressure at a certain temperature.[193] Given that the RH value varies with the temperature, these two quantities are usually measured simultaneously. Moreover, when the RH value approaches 100 %, liquid water formation due to condensation is likely.[194]

As contact measurement, the RH/T (relative humidity/temperature) sensors must be directly implanted at the positions of interests inside an investigated cell. Commercial RH/T sensors e.g., the SHT series from SENSIRION, usually possess a package size of several $mm^3$. They can be well fitted into the flow channels of BPPs for investigating the water contents in the flow fields of LTFCs, with a minimum influence to their regular operations. **Fig. 13** (a) depicts a schematic diagram of the sensor integration. With a sparse array containing 9 evenly distributed RH/T sensors, *J. Zhao et al.* investigated the RH distribution in the cathode flow field of a 50 $cm^2$ single LTFC under different operating conditions, and the results reveal that the down-stream may suffer from water flooding [see also **Fig. 13** (a)], which results to the local CL degradations that is confirmed afterwards by a destructive experiment.[195] Similar studies also addressed commercial scale LTFC single cells and stacks with active areas above 100 $cm^2$.[193,196,197] Assisted with electrochemical measurements, *Z. Zhou et al.* demonstrate the decoupling of the three types of overpotentials in a commercial LTFC stack [197], as shown in **Fig. 13** (b), which provides a direct correlation of local water saturation and efficiency loss.

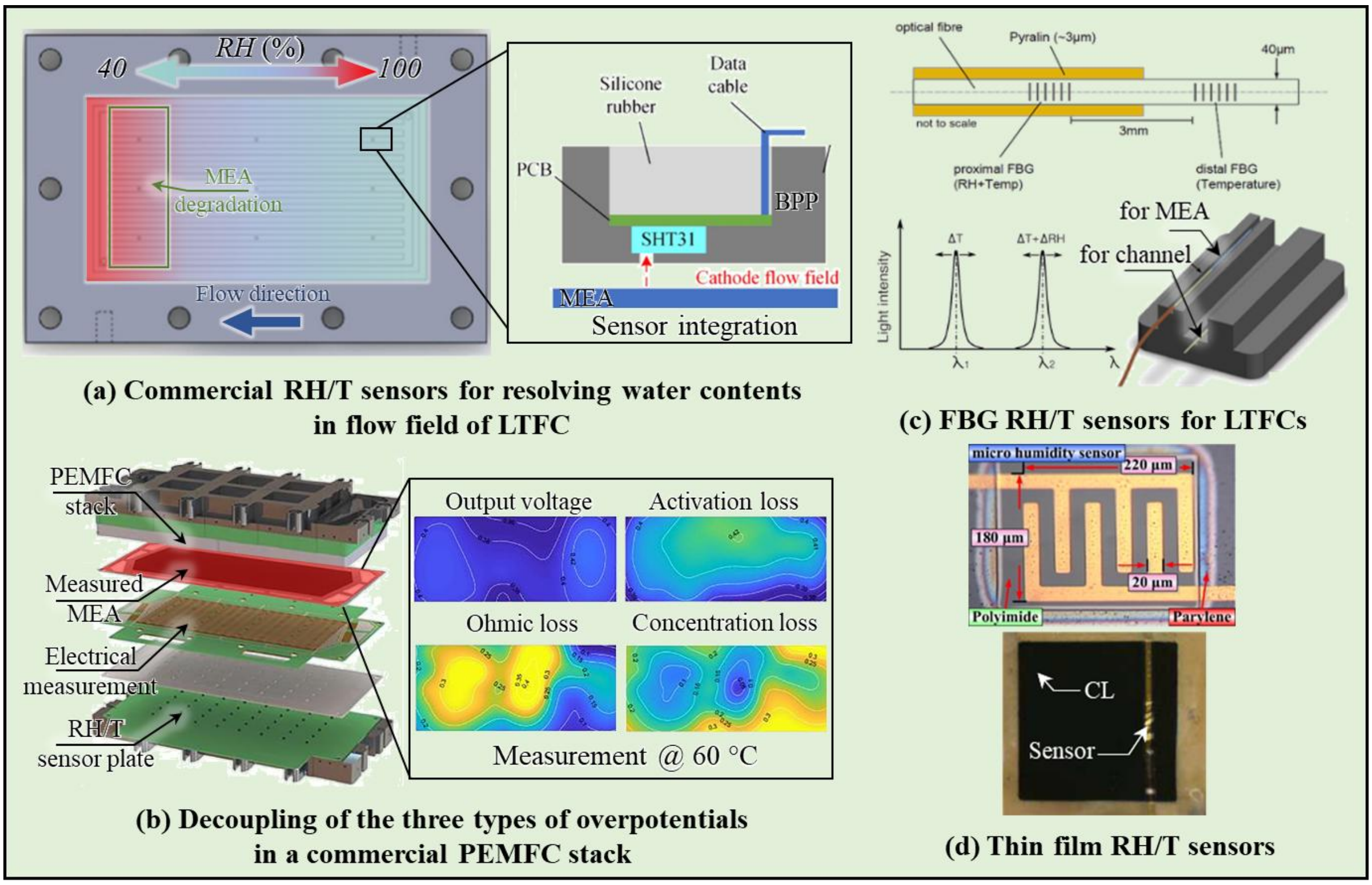

**Fig. 13. Sparse sensor arrays towards mapping gas/liquid transport in LTWEs and LTFCs.** (a) Investigating water contents in the flow field of a LTFC using commercial RH/T sensors. Images adapted from [195]. (b) RH/T and electrochemical co-measurements enable decoupling of localized overpotentials of a commercial PEMFC stack. Images adapted from [197]. (c) The working principle and integration scheme of FBG based RH/T sensors for resolving water contents in LTFC. Images adapted from [194,198]. (d) Left: a capacitive thin film flexible humidity sensor. Right: thin film humidity sensor being hot pressed on the CL into a MEA. Images adapted from [199,200].

In order to investigate the water contents near the MEAs' surfaces in the regions under ribs, RH/T sensors based on optical fiber Bragg grating (FBG) with smaller diameters of approx. 125 μm are introduced [198]. The working principle and integration schemes of using these sensors are depicted in **Fig. 13** (c). In short, the FBG RH/T sensors introduce a periodic variation in the refraction index of the fiber core, generating an environmental sensitive interference peak. Using a single FBG RH/T sensor, *N. David et al.* reveal water accumulations under the ribs in operating LTFCs, which was previously only possible with neutron and X-ray imaging systems.[194,198]. Although the measurements performed near the surface of a MEA are more accurate for estimating the local hydration of the PEM/AEM material, such sensor integration scheme may still perturb the local convective flow and mass transport patterns in an investigated LTWE/LTFC.

Addressing the intrusive risk, flexible thin film RH/T sensors are also developed. Thanks to the micro fabrication technologies, these sensors can be produced at wafer level on thin polymetric foil substrate (approx. 100 μm) with small footprints e.g., < 0.5 mm$^2$ [199], as shown in **Fig. 13** (d). Such thin film sensors not only can be integrated normally in the flow fields but can also hot compressed into a layer of a MEA for measuring water transport [also see **Fig. 13** (d) for the integration scheme].[199,200] *J. Tsujikawa et al.* demonstrate the feasibility of using the latter scheme to investigate the water content at a CL’s surface in a LTFC, which inevitable influences the cell performance due to partial CL coverage.[200]

In addition to the RH/T sensing capabilities, more functionalities, such as flow rate, pressure, hydrogen and oxygen sensing can be integrated into micro sensors.[192,201,202] These multi-functional sensors show potential to enable more comprehensive analysis of the internal conditions in large-scale LTWEs/LTFCs towards optimized flow field designs, MEA configurations as well as combinations of MEAs and flow field patterns, to name a few. However, such minimally intrusive sensor integration schemes, especially the sensors implanted near or directly in contact with a MEA, inevitably influence the regular operations of a LTWE/LTFC under investigation. This implies a critical balance between invasiveness and measurement fidelity, which hinders the implantations of high-density sensor arrays limiting the spatial resolution. In order to avoid the disturbances of sensors to the regular operations of LTWEs/LTFCs, non-invasive sensing schemes that are able to measure the internal conditions of an investigated LTWE/LTFC contactlessly or from outside of the flow fields are required.

### 5.2 Magnetic field sensors and magnetic field imaging

To overcome the intrusive nature of sensors implanted inside LTWEs/LTFCs, developments of contactless analytical tools is required. The use of magnetic fields presents a promising solution for measuring internal electrical current and the transport of conducting fluids without direct physical contact. According to Ampere's circuital law, a magnetic field is generated around an electrical current, as shown in **Fig. 14** (a). The strength of this magnetic field is proportional to the current magnitude and inversely proportional to the distance from the current. For a LTWE/LTFC equipped with non-ferromagnetic BPPs, its external magnetic fields due to the internal current can be collected by multiple magnetic field sensors that are place surrounding the cell in a non-contact manner, as shown in **Fig. 14** (b).[203,204] The 3D distributions of the electrical current inside the investigated LTWE/LTFC can thus be obtained by solving the inverse problem of magnetostatics [see also **Fig. 14** (b)].[204] **Fig. 14** (c) shows an example of magnetic field imaging resolved local current distribution in an operating LTFC, which is in good agreement with the result from a direct segmented cell current measurement. The distribution of electrical current inside a cell directly reflects its electrochemical performance: a uniform current distribution across the active area ensures efficient operations, while local regions of low current density indicate possible bubble accumulation or water flooding, leading to performance losses. Moreover, the magnetic field measurement enables to determine the through plane current that contribute to the electrochemical energy conversion. Such a magnetic field imaging approach is nowadays extensively performed for fault diagnosis of large scale LTFC stacks with active areas above 100 $cm^2$.[204–206]

Recently, this approach is also explored to investigate gas bubbles towards applications in AWEs and AEMWEs. Specifically, given the produced gas bubbles as the only electrically non-conductive phase in an operating AWE/AEMWE, they can thus be identified as local zero current density in reconstructed images.[51] *N. Kumar et al.* demonstrate numerical simulations of magnetic field imaging of gas bubbles in a 2D model, where a pixel resolution of 5 mm was realized, as shown in **Fig. 14** (d) [52]. Such a spatial resolution is however still too coarse to accurately quantify the bubble dynamics, and this is mainly limited by the inverse magnetostatics problem, in which the total number of determined pixels cannot exceed the number of employed magnetic field sensors. Meanwhile, experimental validations of this approach may be limited by the shielding and distortions of magnetic fields due to the presences of ferromagnetic components e.g., Ni PTEs in AWEs/AEMWEs.

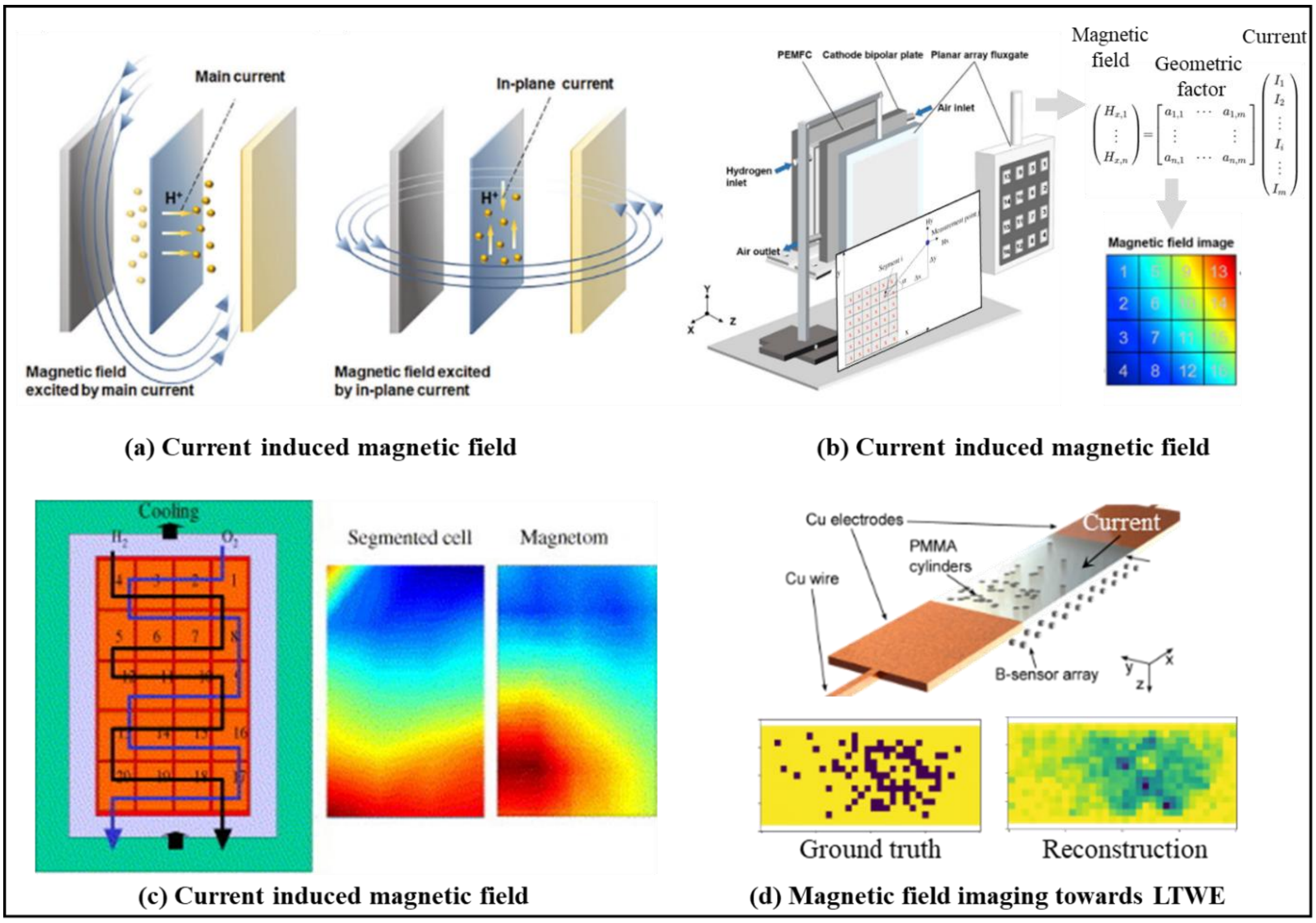

**Fig. 14. Magnetic field imaging of fluid transport in LTWEs and LTFCs.** (a) Schematic demonstration of the electrical current induced magnetic filed in operating LTFCs. Image adapted from [203]. (b) Inverse method for computing local current densities from the measured magnetic field. Images adapted from [203,204]. (c) An example of magnetic field imaging resolved local current distribution in an operating LTFC. Image adapted from [204]. (d) Numerical simulation of magnetic field imaging towards reconstruing gas bubbles in AWEs/AEMWEs. Image adapted from [52].

Finally, it is worth mentioning that the spatial resolution of magnetic field imaging is inherently related to the number of used magnetic field sensors. When the desired number of pixels exceeds the number of sensors, the inversed problem is mathematical ill posed, leading to significant inversion errors.

### 5.3 Summary of sparse sensor array schemes

The sparse sensor array schemes open new horizons for developing innovative analytical systems for investigating larger scale technically relevant LTWE/LTFC stacks, in which alternative physical effects towards sensing the fluid transport related quantities are explored. This strategy enables long-term, real-time monitoring of full-size stacks using affordable electronics.

Despite these advantages, the spatial resolutions of these sensor network schemes are either limited by the intrusive sensor integrations or the inverse problem for image reconstruction. In addition, the existing sensing schemes are mostly able to resolve a single phase e.g., vapor phase for the RH/T sensors [207,208], making them unable to fully characterize the two-phase fluid transport. Future progress can likely be made by advanced signal processing technologies e.g., physics informed neuron network based high fidelity reconstruction. These limitations however highlight the continued need for analytical tools exploring physical effects measuring fluid transport in industrially relevant cells with high spatiotemporal resolution.

## 6. Ultrasonic instrumentations: towards monitoring fluid transport in operating systems

To overcome the limited accessibility and FoV of radiographic imaging systems and the invasiveness of embedded sensor arrays, ultrasonic measurement methods emerge as a promising solution for investigating fluid transport at the device and system levels. Ultrasound, as mechanical waves with frequencies above 20 kHz, are able to deeply penetrate opaque materials, offering sharp contrasts and decent spatiotemporal resolutions in large FoV, using cost-effective instrumentations. These ultrasonic techniques can be roughly classified into passive and active methods. In the context of fluid transport in LTWEs/LTFCs [209], the former collects the ultrasonic (acoustic) wave produced by bubble or droplet evolution events, while the latter explores the interactions of ultrasound and fluid in LTWEs/LTFCs, measuring e.g., the time of flight (ToF) or amplitudes of the resulted signals. To date, a few ultrasonic measurement techniques have been specifically customized to comply with the unique architecture of LTWEs/LTFCs. This section reviews the current developments and future perspectives of studying fluid transport in LTWEs/LTFCs using ultrasonic techniques.

### 6.1 Passive acoustic emission (AE)

Passive acoustic emission (AE) analyse fluid transport in operating LTWEs and LTFCs by acquiring their spontaneously emitted acoustic waves. Specifically, water flows in LTWEs will lead to mechanical oscillations of bubbles at their resonant frequencies, emitting acoustic signals in form of short pulses, as shown in **Fig. 15** (a), and known as the Minnaert resonance.[210] Likewise, bubble detachments and coalescences emit acoustic signals as well.[211] In LTFCs, water evolution may induce structural deformations, such as membrane swelling, and detaching water droplets may impinge on the BPPs, both of which generate acoustic waves. These spontaneously generated AE signals are passively acquired by one or multiple ultrasonic transducers placed out of an investigated LTWE/LTFC. **Fig. 15** (b) plot an exemplary AE signal acquired from an operating LTWE, contain rich information e.g., amplitude, energy and frequency that are related to the fluid transport inside the cell. *M. Maier et al.* reported positive correlation of peak frequency of AE signals and current density of an operating PEMWE, as shown in **Fig. 15** (c). *V. Bethapudi et al.* performed AE measurements for an operating LTFC, in which negative correlations between cumulative AE energy and current density was found, as shown in **Fig. 15** (d) & (e).

However, the passive AE method is unable to localize bubbles and droplets, due to the inability to synchronise bubble/droplet evolution related AE events with signal acquisition, and the lack of knowledge on the exact wave propagation paths inside the complex LTWE/LTFC structures.[212] As a result, the passive AE method can only reveal correlations between signal features, such as amplitude and frequency, and fluid transport-related phenomena in LTWEs and LTFCs, which must be confirmed or interpreted using complementary analytical techniques.[213–217] Therefore, AE serves as a promising *operando* diagnostic or prognostic tool, rather than an analytical method capable of resolving spatiotemporal fluid transport in LTWEs and LTFCs.

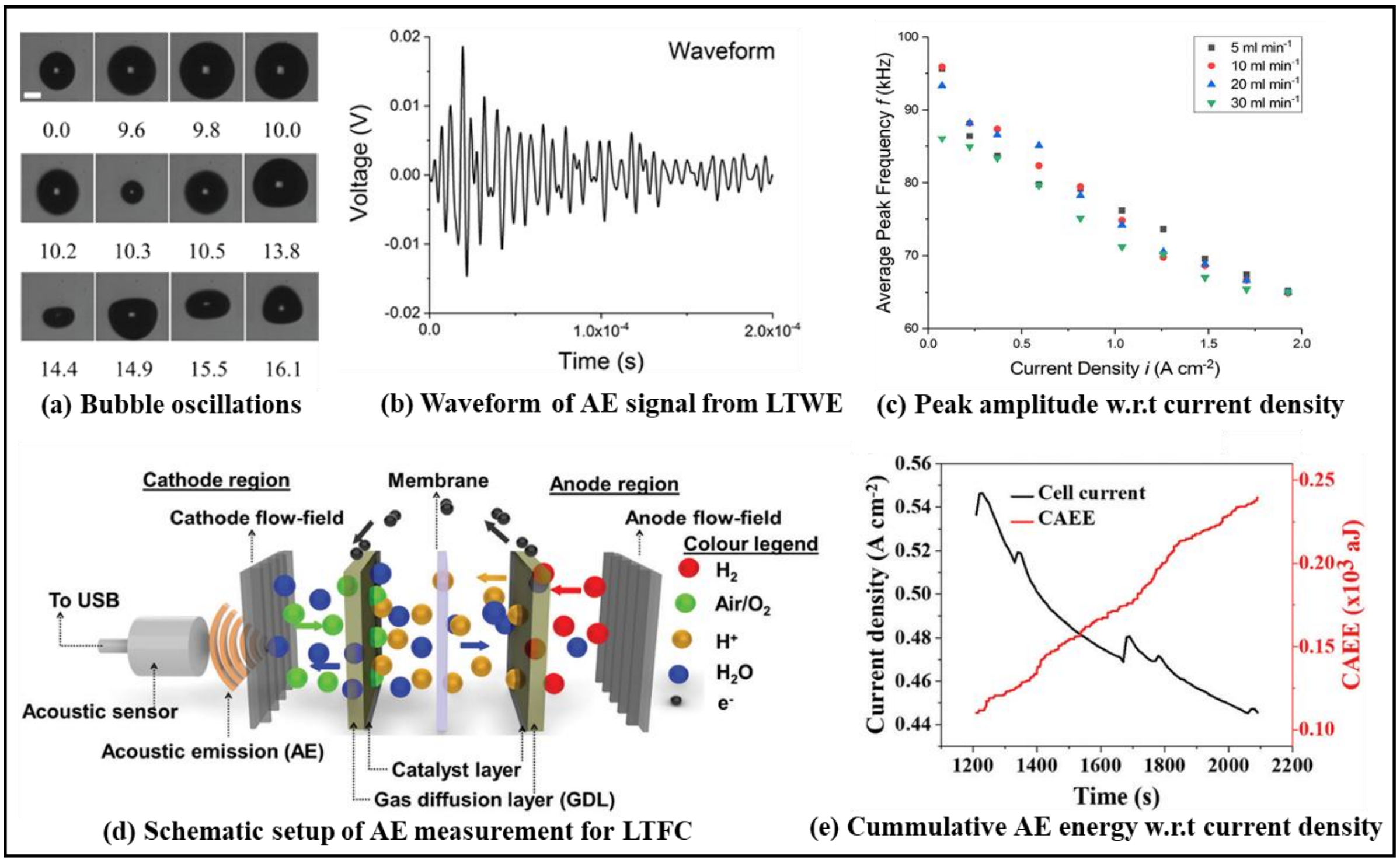

**Fig. 15. Passive acoustic emission (AE) method for LTWEs and LTFCs.** (a) Time-lapses of the bubble oscillations emitting ultrasonic waves. Images adapted from [218]. (b) Typical AE waveform. Image adapted from [214]. (c) Peak frequency of AE signal with respect to current density of investigated LTWE. Image adapted from [214]. (d) Setup of AE measurement for LTFC. Image adapted from [215]. (e) Cumulative AE energy with respect to current density of an investigated LTFC. Image adapted from [215].

## 6.2 Active ultrasonic measurement methods

The inability of passive AE methods to localize fluid transport events can be overcome with active ultrasonic measurement techniques. Among these, ultrasonic imaging is the most widely used, offering direct spatiotemporal visualization of internal fluid structures in operating LTWEs/LTFCs. Additionally, ultrasonic imaging-based velocimetry enables the quantification of convective flows in the cells. This section highlights recent developments on adapting ultrasonic techniques and instrumentations to resolve fluid transport in the GDLs and flow fields of LTWEs and LTFCs, characterized by unique structures and complex geometries, by selecting appropriate ultrasonic wave modes and frequencies. Particular attention is given to the challenges of achieving sufficient resolution and penetration depth, as well as reliable signal interpretation in these complex environments.

### 6.2.1 Ultrasonic imaging and flow velocimetry for LTWEs

*Basic principle of ultrasonic imaging*

**Fig. 16** (a) illustrates the basic principle of ultrasonic pulse-echo imaging, in which short pulses are transmitted into a test cell, and the echoes reflected from its internal structures are detected using the same transducer or transducer array. These echoes arise at interfaces where a mismatch in acoustic impedance presents, a property defined as the product of a material's density and its speed of sound. The time delay between the transmitted and received signals, known as time-of-flight (ToF), provides the depth of the reflecting interface. While, the amplitude of the reflected signal ($r$) is proportional to the acoustic impedance mismatch, enabling the imaging contrast, as given in (6) where *Z1* and *Z2* donate the acoustic impedances at the both sides of an interface [see also **Fig. 15** (a)].

$$r = \left|\frac{Z_1 - Z_2}{Z_1 + Z_2}\right| \tag{6}$$

For a clear perception, **Table III** summarizes the acoustic properties of materials commonly found in LTWEs and LTFCs.

**TABLE III. The Acoustic Properties of the Components in LTWEs**

| **Components/ matters** | **Materials** | **Density ($kg/m^3$)** | **Speed of sound (m/s)** | **Acoustic impedance (Pa·s/m)** |
|---|---|---|---|---|
| Electrolyte | Water | 1000 | 1450 | 1.45 |
| | KOH solution [219] | 1069 | 1580 | 1.69 |
| BPP | Titanium | 4500 | 6070 | 27.32 |
| LGDL | Carbon | 1100 - 1300 | 2350 | 2.59 – 3.06 |
| PTE | Ni [220] | 8900 | 6040 | 53.76 |
| Bubbles | Air | 1.293 | 340 | $4.4 \times 10^{-4}$ |
| | $H_2$ | 0.083 | 1310 | $1.1 \times 10^{-4}$ |
| | $O_2$ | 1.429 | 316 | $4.5 \times 10^{-4}$ |

As shown in **Table III**, water-rich regions enable efficient ultrasound transmission through the BPPs, whereas gas-filled regions reflect almost all the incident waves due to an acoustic impedance mismatch of 6 orders of magnitude between gas and solid. This makes ultrasonic imaging particularly suitable for investigating water-rich environments in LTWEs. Furthermore, gas bubbles generate strong echo signals at gas/liquid interfaces, enabling effective visualizations of gas evolution and transport during LTWE operations.

Nevertheless, it is worth noting that the ultrasonic parameters in imaging applications differ significantly from those in sono-electrolysis studies where ultrasound is applied to actively modulate the bubble dynamics.[221–223] Sono-electrolysis typically employs low frequencies (<1 MHz) continuous ultrasonic waves with high power up to 100 W, which can actively induce cavitation altering bubble dynamics e.g., forced detachments and coalescences. In contrast, ultrasound imaging operates at high frequency (MHz to hundreds of MHz) and low power of $mW/cm^2$ level [224], transmitting short pulses. These conditions ensure that a negligible influence of imaging process on bubble dynamics during water electrolysis.

*Ultrasonic imaging of gas bubbles in flow fields*

For investigating fluid transport within the flow channels e.g., bubbly and slug flows, ultrasound imaging providing spatial resolutions on sub-millimeter level would be sufficient. This is often realized using phased arrays consisting of multiple small ultrasonic transducers that can be individually operated, often below 20 MHz. *M. Maier et al.* reported real-time 3D ultrasoud imaging of an opearting PEMWE with an active surface area of 9 $cm^2$, as shown in **Fig. 16** (b).[225] In their study, a linear (1D) transducer array consisting of 64 elements with a mid-frequency of 5 MHz was used. The array enabled 2D imaging by electronically focusing a subset of elements to form a focusing beam across the imaging plane. Since the (longitudinal) ultrasonic waves were transmitted and received normal to the surface of the PTL, where the signal path was limited to the millimeter-scale flow channel depth, a high 2D imaging update rate of up to 20 kHz was achieved. Full 3D imaging was then achieved by mechanically translating the array across the flow field of the test cell. The echo signals from gas bubbles within the flow channels and the PTL were distinguishable based on their ToF. These echoes were then reconstructed into spatially resolved images that qualitatively represent local gas content, providing a spatial resolution of approx. 1 mm, as shown in **Fig. 16** (b). Due to the strong acoustic impedance mismatch at gas/liquid interfaces, the regions with stronger echo intensities correspond to greater local gas contents.

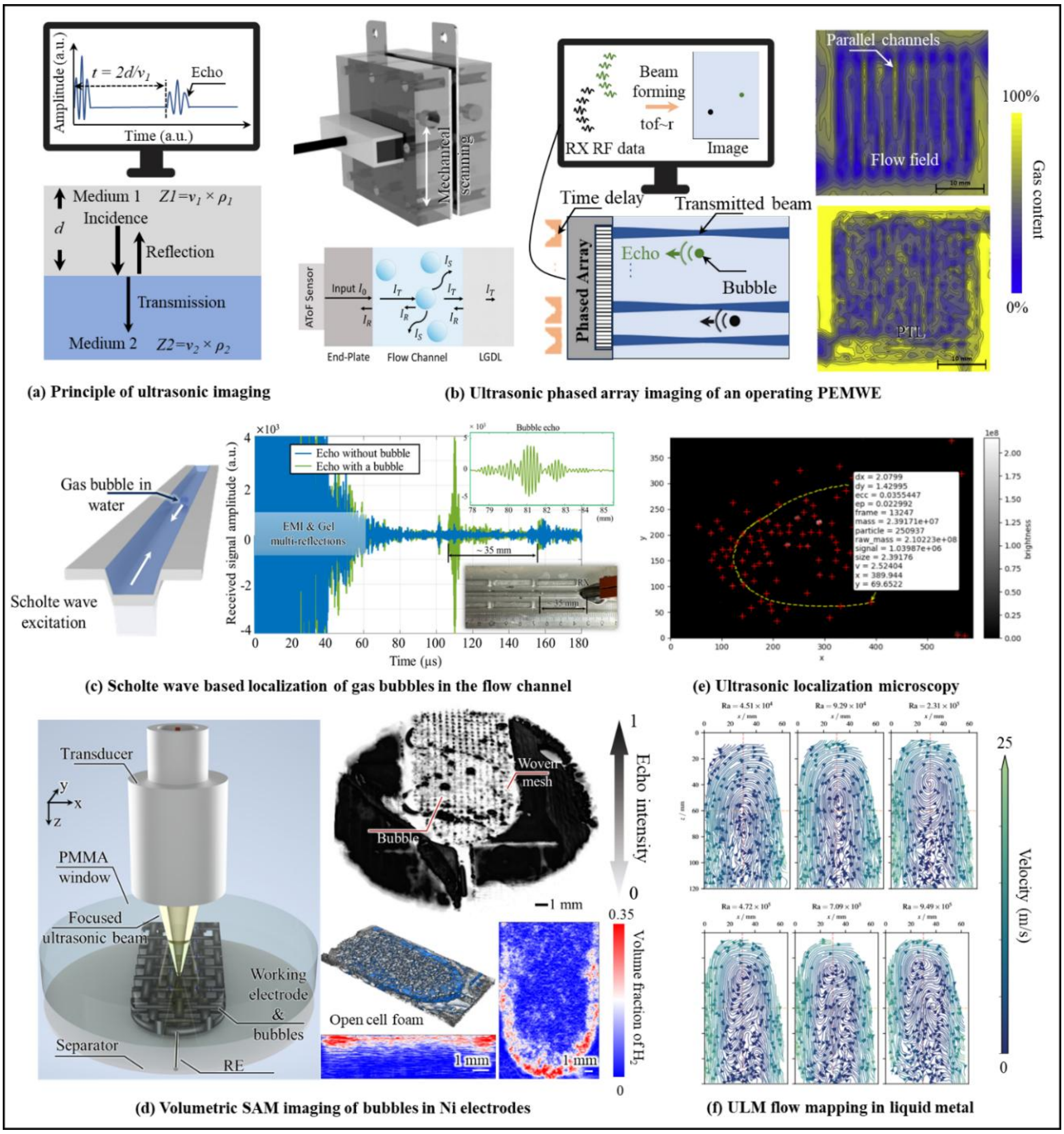

**Fig. 16. Ultrasonic imaging of fluid transport in LTWEs.** (a) Basic principle of pulse-echo mode ultrasonic imaging. (b) Phased array imaging of fluid transport in flow fields of a PWMWE. Image adapted from [225]. (c) Scholte waves based imaging of bubbles in flow channels. Image adapted from [226]. (d) High resolution scanning acoustic microscopy imaging of porous electrodes with woven mesh and open cell foam structures. Image adapted from [219]. (e) The processing flow of PAUDV/ULM. Image adapted from [227]. (f) Demonstration of ULM for flow mapping in liquid metals in a wide range of Reynolds numbers. Images adapted from [228].

To overcome the complexity associated with mechanical scanning of transducer array, ultrasonic guided waves based imaging is adopt with the flow field structures of LTWEs. The key difference between the aforementioned longitudinal and guided waves is that the former propagates perpendicularly to the BPP, while the later propagates in parallel to the BPP. *Z. Dou et al.* proposed a specific guided mode, called Scholte waves, that are able to propagate along the interface of water and BPP was explored, and the principle is shown in **Fig. 16** (c).[226] This in-plane sensing scheme allows the use of a single transducer to investigate fluid transport in each individual flow channel. This method was validated in *ex-situ* experiments, as shown in **Fig. 16** (c) where a syringe induced bubble is localized. Furthermore, this system realized sub-millimeter spatial resolution and a detection limit of gas bubbles less than 100 μm, providing a high frame rate of 2 kHz in a 115 mm long flow channel. Consequently, a single linear transducer array e.g., consisting of 64 elements, can be deployed to monitor the entire flow field of a single cell in commercial scale LTWEs, making the Scholte wave based system more scalable and practical towards system level applications.

*High resolution ultrasound imaging of fluid transport in porous electrodes*

In order to resolve bubble dynamics within the PTL at the pore scale, spatial resolutions at the order of micrometres are required. Achieving such high resolution necessitates the use of high-frequency and tightly focused ultrasound beams. Scanning acoustic microscopy (SAM) employs this principle, using a single-element high-frequency transducer, operating in a frequency range of 50 MHz – 300 MHz, combined with a spherically focused acoustic lens with a large angular aperture. This enables ultrasound to be focused into micrometre-scale regions. Full-field image acquisition is then achieved through mechanical point-by-point scanning. *Z. Dou et al.* demonstrated volumetric SAM imaging of fluid transport in technically relevant Ni electrodes of an operating AWE [219], as shown in **Fig. 16** (d). With a 75 MHz spherically focused transducer, they achieved an axial resolution of 10 μm and a lateral resolution of 70 μm. The study further demonstrated quantitative analysis of gas volume distributions inside electrodes with various configurations under different operating conditions [see **Fig. 16** (d)], establishing a link between cell design, fluid transport, and electrolysis performance. However, due to the inherently time-consuming pointwise raster scanning process, the imaging speed was limited to 180 seconds per volumetric frame for a 900 $mm^2$ area. Real-time imaging is feasible for smaller FoV e.g., 1 s per 5 $mm^2$, or could be achieved by deploying multiple transducers for parallel scanning.

*Ultrasonic velocimetry towards resolving convective*

In addition to structural ultrasound imaging for investigating bubble dynamics, ultrasonic velocimetry, also shows great potential to study convections in the flow channels of LTWEs. Among available techniques, phased array ultrasonic doppler velocimetry (PAUDV) is particularly noteworthy, as it allows for tracking of individual tracer particle's trajectory in complex flows [229], analogue to its optical counterparts, PTV. The typically process is shown in **Fig. 16** (e), involving four steps: ultrasonic imaging is firstly performed using a phased array system; tracer particles are identified based on their motion-induced Doppler shifts or other unique features in echo signals; centroids of individual tracer particles are localized with sub-wavelength precision; the trajectories of particles between consecutive frames are obtained by tracking algorithms.[227] Because the localization precision can surpass the diffraction limit of conventional imaging, this technique is also referred to as super-resolution ultrasonic imaging or ultrasound localization microscopy (ULM).

With this approach, *C. Kupsch et al.* investigated particle suspension flows in the flow channel of a Zin-air battery model [230], in which coated microbubbles that are derived from medical imaging were used as tracer particles, and a lateral resolution of 66 μm corresponding to 1/5 of wavelength was realized. In another study, *D. Weik et al.* explored convections in liquid metals within a wide range of Reynolds numbers, as shown in **Fig. 16** (f), utilizing naturally occurring oxidations as tracer particles.[228] These studies suggest a direct transferability to study convections in LTWEs. However, the influences of tracer particles must be carefully addressed.

In summary, ultrasound imaging and velocimetry open new horizons for understanding of fluid transport induced performance losses in large-scale LTWEs, via non-invasive high spatiotemporal resolution imaging, guiding their optimizations.

### 6.2.2 Ultrasonic methods for water transport in LTFCs

Precise knowledge of liquid water transport in operating LTFCs is crucial for effective water management, avoiding flooding and membrane dehydration, thereby ensuring optimal performance.

*Ultrasonic measurement of water content in GDL*

*Z. Dou et al.* investigated relations of water saturation in GDL and the attenuation and velocity of ultrasonic waves [231], as shown in **Fig. 17** (a). In *ex-situ* experiments using Sigracet 29 BC type GDLs, they found that both attenuation and wave velocity inversely correlate with water saturation. Higher attenuation at lower saturation levels was attributed to strong reflections at gas/solid interfaces [see **Fig. 17** (a)], while variations in wave velocity were related to changes in the effective acoustic properties of the porous GDL medium. Other longitudinal ultrasonic wave based techniques originally developed for Lithium-ion battery researches e.g., resonance spectroscopy for electrode thickness and state of charge measurements, could potentially be adapted to assess the water contents in layered BPP/GDL/BPP structure of LTFCs.[232] However, the accuracy and applicability of these methods depends on precise pre-calibration of ultrasound propagation characteristics, which are highly GDL's material properties and microstructure specific. In other word, further technical efforts are required to generalize this method for a wide range of LTFC designs.

*Ultrasonic measurement of water droplets in flow channels*

For investigating droplets in flow fields of LTFCs, the use of ultrasonic guided waves, particularly the Lamb waves that propagate in plane following the BPPs, offer a promising strategy. Technically, Lamb waves propagate along flat plates, producing mechanical displacements on both sides, as illustrated in inset of **Fig. 17** (b). When lamb waves meet a droplet on plate, the incident waves are partially reflected. This in-plane sensing scheme allows coverage of large areas with significantly fewer transducers, greatly simplifying system design and improving scalability.

*J. Sablowski et al.* reported a Lamb wave tomographic water droplet localization for LTFC scheme [233]. In *ex-situ* experiments, as shown in **Fig. 17** (b), four small piezoelectric disc ultrasonic transducers were mounted to a BPP with an active surface area of approx. 25 $cm^2$, and the Lamb waves were transmitted and received crossing the entire flow field in multiple directions. However, due to the complex wave behavior in the structured BPP geometry, the relationship between signal variation and water droplet presence could not be directly interpreted using physical models. Instead, machine learning was employed for droplet localization. A training dataset was collected by placing droplets at known positions and recording the corresponding signals among the four transducers. This method achieved a localization accuracy of approx. 4 mm for a single 5 μL droplet. This tomographic approach however has limitations, in terms of the detectable droplet size being large for typical commercial flow channels, and transferability of trained model to other real LTFCs. These limitations are primarily attributed to strong scattering at channel-rib boundaries, which distorts the wave path and reduces the signal-to-noise ratio.[234]

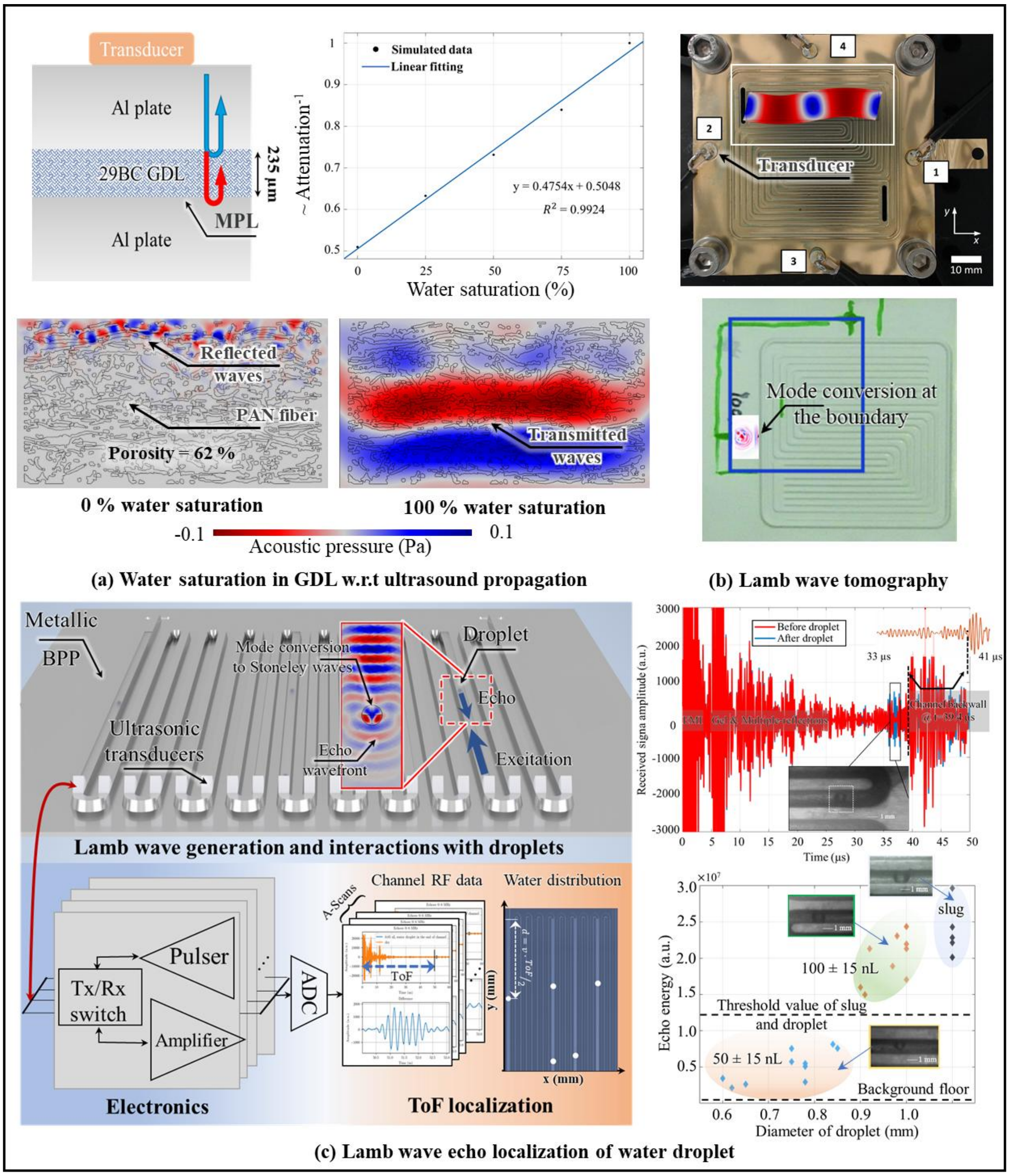

**Fig. 17. Ultrasonic techniques towards investigating water transport in LTFCs.** (a) The investigation of longitudinal ultrasonic wave propagation in GDL: the setup (upper left), the relation of water saturation in GDL and attenuation, reflection of ultrasound in air saturated GDL leading to high attenuation in transmitted waves (lower left), and ultrasound propagation in water saturated GDL (lower right). Images adapted from [231]. (b) Lamb wave tomography for localizing water droplet scheme: the setup, inset shows a cross-sectional view of Lamb wave propagation (upper), and scattering of Lamb waves at channel/rib boundaries (lower). Images adapted from [233], [234]. (c) Lamb wave echo localization of water droplets: the working principle (left), and the experimental demonstrations (right). Images adapted from [235].

To overcome these challenges, *Z. Dou et al.* proposed a Lamb-wave echo-localization based water monitoring scheme, as illustrated in **Fig. 17** (c) [235]. In this scheme, Lamb waves are selectively generated along individual flow channels where water droplets are most likely to accumulate, using multiple ultrasonic transducers. This allows each free-standing flow channel to be approximated as a simple flat plate, which significantly improves wave transmission efficiency and simplifies the signal interpretation. In *ex-situ* experiments, the system accurately localized droplets as small as 50 nL across the entire flow field covering an area of 25 $cm^2$, providing a spatial resolution of 1.3 mm at a frame rate of 2 kHz. Such a sensitivity and resolution are sufficient for determining water accumulation and flow regimes in the flow channel, validating the feasibility of high-resolution, real-time droplet tracking in practical LTFC geometries.

Despite these advances, the Lamb wave based measurement systems have so far only been demonstrated in *ex-situ* experiments. To make this method a mature research instrumentation for *operando* applications, compact ultrasonic transducers integrated into LTFCs would be necessary.[235]

As a summary, the key advantage of these active ultrasonic techniques is their ability of real-time monitoring of fluid transport in large flow fields, providing decent spatial resolutions.

### 6.3 Summary

This section reviews the current developments of ultrasonic measurement techniques for liquid transport in LTWEs/LTFCs. A comparative overview of these techniques is provided in **Table IV**.

**Table IV. Summary of ultrasound measurement techniques for fluid transport in LTWEs/LTFCs**

| Method | Wave mode | Application | Spatial Resolution | Temporal Resolution | Advantages | Other comments |
|---|---|---|---|---|---|---|
| **Passive AE** | N.A. | Diagnostic/ prognostic | N.A. | Real-time | Simple instrumentation | No localization possible; indirect interpretation |
| **Pulse-echo phased array imaging** | Longitudinal | Bubbles in flow channels of LTWEs | ~1 mm | Up to 20 kHz | Simple instrumentation | Scanning required |
| | Scholte / Lamb | Bubbles in flow channels of LTWEs/droplet in LTFCs | ~ 1 mm | 2 kHz | Good scalability; integration possible | Wave mode complexity; *in-situ* not yet reported |
| **SAM** | Longitudinal | PTL microstructure (LTWEs, AWE) | ~ µm | 180 s/frame [219]; Real-time possible | Micrometer level resolution; pore-scale insights | Slow raster scan; commercial SAM available |
| **Spectroscopy** | Longitudinal | Water saturation (LTFC GDL) | Sensor size determined | N.A. | Sensitive to saturation and interfaces | Requires calibration; complex signal interpretation |
| **ULM** | Harmonic Doppler | Convection in LTWEs | ~ 60 µm (sub diffraction limit) | Real-time possible | Super-resolution tracking | Requires tracer particles |

As a final remark, ultrasonic instrumentations present a significant step toward developing scalable, cost-effective, and non-invasive system-level monitoring tools, bridging the gap between fundamental research on small prototypes and practical implementations. They demonstrate the potential to replace the costly and complicated radiographic systems for regular laboratory studies.

Despite these advantages, these ultrasonic techniques require further developments towards wide *operando* adoptions. Meanwhile, other proven ultrasonic methods used in other fields e.g., medical, batteries and non-destructive testing, can be actively adapted for LTWE/LTFCs with suitable instrumentations and integration strategies.

## 7. Future prospects

The advancement of green hydrogen technologies significantly relies on improved understanding and control of complex fluid transport in practical low-temperature water electrolyzers (LTWEs) and fuel cells (LTFCs). This review systematically summarized the fluid transport challenges, and how analytical technologies reveal the relations of these transport processes and performance loss, contributing to the optimizations.

In summary, conventional optical and radiographic imaging have significantly enriched the insights into the optimizations of LTWEs and LTFCs in small-scale lab prototype studies. However, these techniques face fundamental limitations in terms of scalability, hindering the transfer from lab prototype to industrial applications. This gap has created a pressing need for innovations in analytical instrumentations that can investigate cells at larger scales and under real operating conditions, Embedded sensor networks and ultrasonic techniques represent promising research directions, extending *operando* investigations towards industrial-scale cells. However, no single method is yet capable of real-time capturing the multiphase and multiscale fluid transport in practical LTWEs and LTFCs. Instead, each technique contributes complementary strengths in corresponding applications are summarized in **Table V** and **Fig. 18**.

The perspectives of fluid transport analysis in LTWEs and LTFCs are below.

(1) The maturation of the scalable and effective measurement systems based on ultrasound and minimized sensors are essential for accelerating general laboratory and online research outputs. By making real-time monitoring at device and system levels technologically accessible, the optimization cycles for industrially relevant components and systems will be significantly accelerated, driving cost-effective green hydrogen.

(2) Enriching knowledge on the practical fluid transport issues does not rely on replacing current methods with novel systems, but rather to combine suitable complementary tools for multiscale investigations. Synergistic combinations of pore scale and device level investigations e.g., coupling the insights obtained by scanning acoustic microscopy imaging and ultrasonic guided waves based flow field monitoring, will provide a more complete picture of fluid transport, linking fundamental mechanisms to practical optimizations.

(3) From an instrumentation engineering perspective, it is believed that, future breakthrough will also be made by a paradigm shift to computational imaging, where advanced signal processing frameworks e.g., compressive sensing, ghost imaging, and physics-informed neural networks (PINNs) can greatly enhance the scalability and fidelity of current instrumentations.[236] For instance, with these methods, the contrast issue of X-ray and neutron imaging could be greatly improved, while reducing the required beam brightness. Likewise, the use of PINNs would drastically improve the spatial resolutions using only a few sparsely distributed e.g., magnetic or ultrasonic sensors.

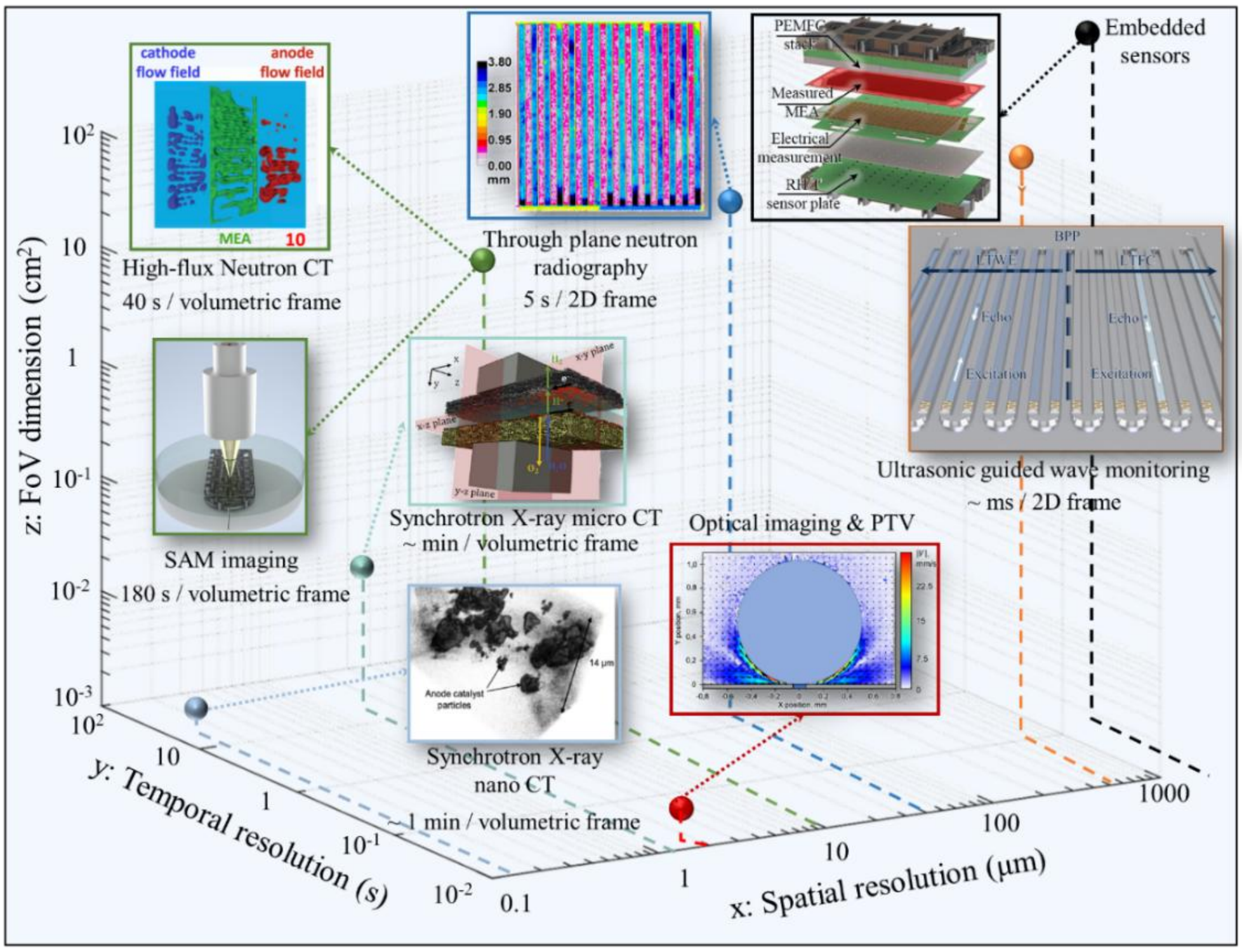

**Fig. 18. A summary of performance and applications of modern analytical technologies.** Images adapted from [88,95,160–162,219,226,235].

**Table V. Summary of state-of-the-art analytical methods for fluid transport in LTWEs/LTFCs**

| Analytical Technique | Target of Investigation | Key Features (e.g., Resolution) | Advantages | Limitations |
|---|---|---|---|---|
| **Optical Imaging** | Liquid water and bubble flow | High spatial and temporal resolution | Relatively simple and low cost | Requires transparent components; limited penetration depth in GDLs/PTLs |
| **X-ray Radiography/ Tomography** | Two-phase flow, water distribution, material defects | High spatial resolution (μm level) | Non-invasive; deep penetration; can be used operando | High cost and limited access; requires dedicated facilities (e.g., synchrotron); poor contrast in LTWEs |
| **Embedded Micro-sensors** | Localized pressure, temperature, or water content | High temporal resolution; direct quantitative data | Provides localized, real-time data from within the cell | Invasive; sensor placement can affect flow; complex integration |
| **Neutron Imaging/ Tomography** | Water content distribution in cells with metal components | High sensitivity to hydrogen/water | Non-invasive; deep penetration, complements X-rays | Very high cost; limited access to neutron sources; lower spatial resolution than using X-rays |
| **Ultrasonic Imaging** | Liquid water content; | High spatial resolution (μm) | Non-invasive; relatively low cost; fast imaging speed | Signal attenuation can be an issue; complex data interpretation |

As a final remark, overcoming the fluid transport challenge in LTWEs and LTFCs requires close cooperation between electrochemical and instrumentation communities, towards rational device design, system operation and integration, which would ultimately facilitate scale-up of reliable and efficient green hydrogen technologies.

## Acknowledgements

The authors would like to thank C. Othamani for valuable discussions. This work was funded by the German Aerospace Center Project Sponsor (DLR-PT) within the support program "Industrial Joint Research" (IGF) of the German Federal Ministry of Economics and Technology under Grant 1F23280N, the Deutsche Forschungsgemeinschaft (DFG) under grant 512483557 (BU2241/9-1), and the Helmholtz Association Innovation pool project "Solar Hydrogen" and the Hydrogen Lab of the School of Engineering of TU Dresden.

## Declaration of interests

The authors declare no competing interests.

## Author contributions

Z. D. conceptualization, investigation, formal analysis, writing – original draft, and writing – review & editing. L. T. formal analysis, and writing – original draft, and writing – review & editing. T. L. investigation, writing – review & editing. H. R. formal analysis, writing – review & editing. X. Y. formal analysis, writing – review & editing. L. B. investigation, formal analysis, and writing – review & editing. D. W. investigation, formal analysis, writing – original draft, and writing – review & editing. H. H. investigation, formal analysis, writing – review & editing, project administration, and funding acquisition. K. E. Writing – review & editing, supervision, project administration, funding acquisition. J. C. Writing – review & editing, supervision, project administration, and funding acquisition.

## Nomenclature

| **Abbreviations** | **Meanings** |
|---|---|
| 1D | One-dimensional |
| 2D | Two-dimensional |
| 3D | Three-dimensional |
| AEM | Anion exchange membrane |
| AWE | Alkaline water electrolyzer |
| Al | Aluminium |
| BPP | Bipolar plate |
| CCM | Catalysts coated membrane |
| CT | Computational tomography |
| CV | Cyclic voltammetry |
| EOD | Electro-osmotic drag |
| EIS | Electrochemical impedance spectroscopy |
| FoV | Field of view |
| GDL | Gas diffusion layer |
| $H_2$ | Hydrogen |
| HER | Hydrogen evolution reaction |
| HOR | Hydrogen oxidation reaction |
| KOH | Potassium hydroxide |
| LGDL | Liquid gas diffusion layer |
| LTFC | Low temperature fuel cell |
| LTWE | Low temperature water electrolyzer |
| Ni | Nickel |
| MEA | Membrane electrode assembly |
| Ni | Nickel |
| $O_2$ | Oxygen |
| OER | Oxygen evolution reaction |
| ORR | Oxygen reduction reaction |
| PEM | Proton exchange membrane |
| PEMFC | Proton exchange membrane fuel cell |
| SAM | Scanning acoustic microscopy |
| SNR | Signal to noise ratio |
| RH | Relative humidity |
| ToF | Time of flight |
| Ti | Titanium |

## Reference

1. The Guardian. 'Era of global boiling has arrived,' says UN chief as July set to be hottest month on record [Internet]. 2023. Available from: https://www.theguardian.com/science/2023/jul/27/scientists-july-world-hottest-month-record-climate-temperatures#:~:text=The era of global warming,the beginning%2C" Guterres said.
2. Panchenko VA, Daus Y V., Kovalev AA, Yudaev I V., Litti Y V. Prospects for the production of green hydrogen: Review of countries with high potential. Int J Hydrogen Energy [Internet]. 2023;48(12):4551–71.
3. Abbasi T, Abbasi SA. "Renewable" hydrogen: Prospects and challenges. Renew Sustain Energy Rev. 2011;15(6):3034–40.
4. Oliveira AM, Beswick RR, Yan Y. A green hydrogen economy for a renewable energy society. Curr Opin Chem Eng [Internet]. 2021;33:100701. Available from: https://doi.org/10.1016/j.coche.2021.100701
5. Incer-Valverde J, Korayem A, Tsatsaronis G, Morosuk T. "Colors" of hydrogen: Definitions and carbon intensity. Energy Convers Manag. 2023;291(May):117294.
6. Hassan Q, Algburi S, Sameen AZ, Salman HM, Jaszczur M. Green hydrogen: A pathway to a sustainable energy future. Int J Hydrogen Energy. 2024;50:310–33.
7. XPS Extreme Power System [Internet]. Available from: https://infinityfuel.com/solutions/fuel-cell-power/
8. Burke KA. Fuel cells for space science applications. In: 1st International Energy Conversion Engineering Conference IECEC. 2003.
9. Cleghorn S. How membrane technology is driving the commercialization of fuel cells in the automotive industry [Internet]. Available from: https://hydrogencouncil.com/en/how-membrane-technology-is-driving-the-commercialization-of-fuel-cells-in-the-automotive-industry/
10. What does it take to create a hydrogen home [Internet]. Available from: www.hydrogenfuelnews.com/hydrogen-home-energy/8551340/?utm_content=cmp-true
11. Global Hydrogen Review 2025. The International Energy Agency. 2025. Available from: https://www.iea.org/reports/global-hydrogen-review-2025
12. Hydrogen Shot. Available from: https://www.energy.gov/eere/fuelcells/hydrogen-shot
13. Santos AL, Cebola MJ, Santos DMF. Towards the hydrogen economy—a review of the parameters that influence the efficiency of alkaline water electrolyzers. Energies. 2021;14(11).
14. Wang Y, Pang Y, Xu H, Martinez A, Chen KS. PEM Fuel cell and electrolysis cell technologies and hydrogen infrastructure development - a review. Energy Environ Sci. 2022;15(6):2288–328.
15. Makhsoos A, Kandidayeni M, Pollet BG, Boulon L. A perspective on increasing the efficiency of proton exchange membrane water electrolyzers– a review. Int J Hydrogen Energy. 2023;48(41):15341–70.
16. Vidas L, Castro R. Recent developments on hydrogen production technologies: State-of-the-art review with a focus on green-electrolysis. Appl Sci. 2021;11(23).
17. Rashid MM, Mesfer MK Al, Naseem H, Danish M. Hydrogen Production by Water Electrolysis: A Review of Alkaline Water Electrolysis, PEM Water Electrolysis and High Temperature Water Electrolysis. Int J Eng Adv Technol. 2015;(3):2249–8958.
18. Miller HA, Bouzek K, Hnat J, Loos S, Bernäcker CI, Weißgärber T, et al. Green hydrogen from anion exchange membrane water electrolysis: A review of recent developments in critical materials and operating conditions. Sustain Energy Fuels. 2020;4(5):2114–33.
19. Liu RT, Xu ZL, Li FM, Chen FY, Yu JY, Yan Y, et al. Recent advances in proton exchange membrane water electrolysis. Chem Soc Rev. 2023;52(16):5652–83.
20. Fujiwara N, Nagase H, Tada S, Kikuchi R. Hydrogen Production by Steam Electrolysis in Solid Acid Electrolysis Cells. ChemSusChem. 2021;14(1):417–27.
21. Zucconi A, Hack J, Stocker R, Suter TAM, Rettie AJE, Brett DJL. Challenges and opportunities for characterisation of high-temperature polymer electrolyte membrane fuel cells: a review. J Mater Chem A. 2024;12(14):8014–64.
22. Green hydrogen production. Available from: www.siemens-energy.com/global/en/offerings/renewable-

energy/hydrogen-solutions.html?gclid=CjwKCAjwxOymBhAFEiwAnodBLPfK6pscSrr6f6PJ5_m63Jj9ImNhs9fnLvQXg9RUmu7hvXkWjij8XhoCzsAQAvD_BwE
23. Yu S, Li K, Wang W, Xie Z, Ding L, Kang Z, et al. Tuning Catalyst Activation and Utilization Via Controlled Electrode Patterning for Low-Loading and High-Efficiency Water Electrolyzers. Small. 2022;18(14).
24. Wang W, Yu S, Li K, Ding L, Xie Z, Li Y, et al. Insights into the rapid two-phase transport dynamics in different structured porous transport layers of water electrolyzers through high-speed visualization. J Power Sources. 2021;516(January):230641.
25. Majasan JO, Cho JIS, Dedigama I, Tsaoulidis D, Shearing P, Brett DJL. Two-phase flow behaviour and performance of polymer electrolyte membrane electrolysers: Electrochemical and optical characterisation. Int J Hydrogen Energy. 2018;43(33):15659–72.
26. Alia SM, Reeves KS, Baxter JS, Cullen DA. The Impact of Ink and Spray Variables on Catalyst Layer Properties, Electrolyzer Performance, and Electrolyzer Durability. J Electrochem Soc. 2020;167(14):144512.
27. Ji M, Wei Z. A review of water management in polymer electrolyte membrane fuel cells. Energies. 2009;2(4):1057–106.
28. Alavijeh AS, Khorasany RMH, Nunn Z, Habisch A, Lauritzen M, Rogers E, et al. Microstructural and Mechanical Characterization of Catalyst Coated Membranes Subjected to In Situ Hygrothermal Fatigue. J Electrochem Soc. 2015;162(14):F1461–9.
29. Lu Z, Zhu W, Yu X, Zhang H, Li Y, Sun X, et al. Ultrahigh hydrogen evolution performance of under-water "superaerophobic" MoS2 nanostructured electrodes. Adv Mater. 2014;26(17):2683–7.
30. Yang WJ, Wang HY, Lee DH, Kim YB. Channel geometry optimization of a polymer electrolyte membrane fuel cell using genetic algorithm. Appl Energy. 2015;146:1–10.
31. Zhang X, Zhou W, Huang Y, Xie L, Li T, Kang H, et al. Pulsed dynamic electrolysis enhanced PEMWE hydrogen production: Revealing the effects of pulsed electric fields on protons mass transport and hydrogen bubble escape. J Energy Chem. 2025;100:201–14.
32. Engelhardt V. Die Elektrolyse des Wassers: Ihre Durchführung und Anwendung (Classic Reprint). London: Forgotten Books; 2018. 394 p.
33. SUNFIRE RECEIVES PURCHASE ORDER FOR 100 MW PRESSURIZED ALKALINE ELECTROLYZER. Available from: https://www.sunfire.de/en/news/detail/sunfire-receives-purchase-order-for-100-mw-pressurized-alkaline-electrolyzer
34. The WESTKÜSTE100 real-world laboratory in detail [Internet]. Available from: https://www.westkueste100.de/en/project-description/
35. REFHYNE project overview [Internet]. Available from: https://www.refhyne.eu/about/
36. Xu Q, Zhang L, Zhang J, Wang J, Hu Y, Jiang H, et al. Anion Exchange Membrane Water Electrolyzer: Electrode Design, Lab-Scaled Testing System and Performance Evaluation. EnergyChem. 2022;4(5):100087.
37. Liu Q, Lan F, Chen J, Zeng C, Wang J. A review of proton exchange membrane fuel cell water management: Membrane electrode assembly. J Power Sources. 2022;517(June 2021):230723.
38. Li H, Tang Y, Wang Z, Shi Z, Wu S, Song D, et al. A review of water flooding issues in the proton exchange membrane fuel cell. J Power Sources. 2008;178(1):103–17.
39. Xu S, Liao P, Yang D, Li Z, Li B, Ming P, et al. Liquid water transport in gas flow channels of PEMFCs: A review on numerical simulations and visualization experiments. Int J Hydrogen Energy. 2023;48(27):10118–43.
40. Miura K. Stiffness reduction and collagenase resistance of aging lungs measured using scanning acoustic microscopy. PLoS One. 2022;17(2 February):1–17.
41. Chaudhari NK, Jin H, Kim B, Lee K. Nanostructured materials on 3D nickel foam as electrocatalysts for water splitting. Nanoscale. 2017;9(34):12231–47.
42. Kandlikar SG, Garofalo ML, Lu Z. Water management in a PEMFC: Water transport mechanism and material degradation in gas diffusion layers. Fuel Cells. 2011;11(6):814–23.
43. Ous T, Arcoumanis C. Degradation aspects of water formation and transport in Proton Exchange Membrane Fuel Cell: A review. J Power Sources. 2013;240:558–82.
44. Monzon LMA, Coey JMD. Magnetic fields in electrochemistry: The Lorentz force. A mini-review. Electrochem commun. 2014;42:38–41.
45. Wu J, Zi Yuan X, Wang H, Blanco M, Martin JJ, Zhang J. Diagnostic tools in PEM fuel cell research: Part II. Physical/chemical methods. Int J Hydrogen Energy. 2008;33(6):1747–57.
46. Maier M, Smith K, Dodwell J, Hinds G, Shearing PR, Brett DJL. Mass transport in PEM water electrolysers: A review. Int J Hydrogen Energy. 2022;47(1):30–56.
47. Tang Z, Huang QA, Wang YJ, Zhang F, Li W, Li A, et al. Recent progress in the use of electrochemical impedance spectroscopy for the measurement, monitoring, diagnosis and optimization of proton exchange membrane fuel cell performance. J Power Sources. 2020;468:228361.
48. Hernández-Gómez Á, Ramirez V, Guilbert D. Investigation of PEM electrolyzer modeling: Electrical domain, efficiency, and specific energy consumption. Int J Hydrogen Energy. 2020;45(29):14625–39.
49. Yuan S, Zhao C, Cai X, An L, Shen S, Yan X, et al. Bubble evolution and transport in PEM water electrolysis: Mechanism, impact, and management. Prog Energy Combust Sci. 2023;96(August 2022).
50. Ijaodola OS, El- Hassan Z, Ogungbemi E, Khatib FN, Wilberforce T, Thompson J, et al. Energy efficiency improvements by investigating the water flooding management on proton exchange membrane fuel cell (PEMFC). Energy. 2019;179:246–67.
51. Krause L, Kumar N, Wondrak T, Gumhold S, Eckert S, Eckert K. Current Tomography - Localization of void fractions in conducting liquids by measuring the induced magnetic flux density. 2023;1–6.
52. Kumar N, Krause L, Wondrak T, Eckert S, Eckert K, Gumhold S. Robust Reconstruction of the Void Fraction from Noisy Magnetic Flux Density Using Invertible Neural Networks. Sensors. 2024;24(4):1–29.
53. Nafion$^{TM}$ Membranes, Dispersions, and Resins [Internet]. Available from: https://www.nafion.com/en/products
54. Dunwoody D, Leddy J. Proton exchange membranes: The view forward and back. Electrochem Soc Interface. 2005;14(3):37–9.
55. Xu B, Ouyang T, Wang Y, Yang Y, Li J, Jiang L, et al. Progresses on two-phase modeling of proton exchange membrane water electrolyzer. Energy Rev. 2024;3(3):100073.
56. Rocha C, Knöri T, Ribeirinha P, Gazdzicki P. A review on flow field design for proton exchange membrane fuel cells: Challenges to increase the active area for MW applications. Renew Sustain Energy Rev. 2024;192(October 2023).
57. Kerkoub Y, Benzaoui A, Haddad F, Ziari YK. Channel to rib width ratio influence with various flow field designs on performance of PEM fuel cell. Energy Convers Manag. 2018;174(August):260–75.
58. Zhou H, Chen B, Meng K, Luo M, Li P, Tu Z. Combination effect of flow channel configuration and anode GDL porosity on mass transfer and performance of PEM water electrolyzers. Sustain Energy Fuels. 2022;6(17):3944–60.
59. Zagoraiou E, Krishan S, Siriwardana A, Moschovi AM, Yakoumis I. Performance of Stainless-Steel Bipolar Plates (SS-BPPs) in Polymer Electrolyte Membrane Water Electrolyser (PEMWE): A Comprehensive Review. Compounds. 2024;4(2):252–67.
60. Tang A, Crisci L, Bonville L, Jankovic J. An overview of bipolar plates in proton exchange membrane fuel cells. J Renew Sustain Energy. 2021;13(2).
61. Karimi S, Fraser N, Roberts B, Foulkes FR. A review of metallic bipolar plates for proton exchange membrane fuel cells: Materials and fabrication methods. Adv Mater Sci Eng. 2012;2012.
62. Jeong K Il, Oh J, Song SA, Lee D, Lee DG, Kim SS. A review of composite bipolar plates in proton exchange membrane fuel cells: Electrical properties and gas permeability. Compos Struct. 2021;262(November 2020):113617.
63. Okonkwo PC, Emori W, Uzoma PC, Mansir IB, Radwan AB, Ige OO, et al. A review of bipolar plates materials and graphene coating degradation mechanism in proton exchange membrane fuel cell. Int J Energy Res. 2022;46(4):3766–81.
64. Tanaka S, Nagumo K, Yamamoto M, Chiba H, Yoshida K, Okano R. Fuel cell system for Honda CLARITY fuel cell. eTransportation. 2020;3:100046.
65. Selamet, Ömer F., M. Caner Acar, Mahmut MDM and YK. Effects of operating parameters on the performance of a high-pressure proton exchange membrane electrolyzer. Int J Energy Res. 2013;37(5):457–67.
66. Dana metallic BPPsDana develops Metallic Bipolar Plates for Hydrogen PEM Electrolyzers [Internet]. Available from: https://etn.news/buzz/dana-metallic-bipolar-plates-pem-electrolyzers-hydrogen-details
67. Graphite Bipolar Plates [Internet]. Available from: https://www.schunk-group.com/carbon-technology/en/products/bipolar-plates
68. Lorenz O, Kühne A, Rudolph M, Diyatmika W, Prager A, Gerlach JW, et al. Role of reaction intermediate diffusion on the performance of platinum electrodes in solid acid fuel cells. Catalysts. 2021;11(9).
69. Yang F, Kim MJ, Brown M, Wiley BJ. Alkaline Water Electrolysis at 25 A cm−2 with a Microfibrous Flow-through Electrode. Adv Energy Mater. 2020;10(25).
70. El-Kharouf A, Mason TJ, Brett DJL, Pollet BG. Ex-situ characterisation of gas diffusion layers for proton exchange membrane fuel cells. J Power Sources. 2012;218:393–404.
71. Omrani R, Shabani B. Gas diffusion layers in fuel cells and electrolysers: A novel semi-empirical model to predict electrical conductivity of sintered metal fibres. Energies. 2019;12(5):1–17.
72. Yang H, Driess M, Menezes PW. Self-Supported Electrocatalysts for Practical Water Electrolysis. Adv Energy Mater. 2021;11(39).
73. Eon Chae J, Choi J, Lee S, Park C, Kim S. Effects of fabrication parameters of membrane–electrode assembly for high-performance anion exchange membrane fuel cells. J Ind Eng Chem. 2024;133(August 2023):255–62.

74. Hyun Oh J, Ho Han G, Kim J, Eun Lee J, Kim H, Kyung Kang S, et al. Self-supported electrodes to enhance mass transfer for high-performance anion exchange membrane water electrolyzer. Chem Eng J. 2023;460(January):141727.
75. Bühler M, Holzapfel P, McLaughlin D, Thiele S. From Catalyst Coated Membranes to Porous Transport Electrode Based Configurations in PEM Water Electrolyzers. J Electrochem Soc. 2019;166(14):F1070–8.
76. Yuan XZ, Shaigan N, Song C, Aujla M, Neburchilov V, Kwan JTH, et al. The porous transport layer in proton exchange membrane water electrolysis: perspectives on a complex component. Sustain Energy Fuels. 2022;6(8):1824–53.
77. Athanasaki G, Jayakumar A, Kannan AM. Gas diffusion layers for PEM fuel cells: Materials, properties and manufacturing – A review. Int J Hydrogen Energy. 2023;48(6):2294–313.
78. Truong VM, Duong NB, Yang H. Effect of gas diffusion layer thickness on the performance of anion exchange membrane fuel cells. Processes. 2021;9(4):1–10.
79. Zhang J, Zhu W, Huang T, Zheng C, Pei Y, Shen G, et al. Recent Insights on Catalyst Layers for Anion Exchange Membrane Fuel Cells. Adv Sci. 2021;8(15):1–26.
80. Zhang T, Meng L, Chen C, Du L, Wang N, Xing L, et al. Similarities and Differences between Gas Diffusion Layers Used in Proton Exchange Membrane Fuel Cell and Water Electrolysis for Material and Mass Transport. Adv Sci. 2024;2309440:1–21.
81. He Y, Cui Y, Zhao Z, Chen Y, Shang W, Tan P. Strategies for bubble removal in electrochemical systems. Energy Rev. 2023;2(1).
82. Taqieddin A, Allshouse MR, Alshawabkeh AN. Editors' Choice—Critical Review—Mathematical Formulations of Electrochemically Gas-Evolving Systems. J Electrochem Soc. 2018;165(13):E694–711.
83. Yang X, Karnbach F, Uhlemann M, Odenbach S, Eckert K. Dynamics of Single Hydrogen Bubbles at a Platinum Microelectrode. Langmuir. 2015;31(29):8184–93.
84. Fritz W. Maximum Volume of Vapor Bubbles. Phys Zeitschr. 1935;36:379–384.
85. Demirkır Ç, Wood JA, Lohse D, Krug D. Life beyond Fritz: On the detachment of electrolytic bubbles. Langmuir. 2024;40(39):20474–20484.
86. Yang X, Baczyzmalski D, Cierpka C, Mutschke G, Eckert K. Marangoni convection at electrogenerated hydrogen bubbles. Phys Chem Chem Phys. 2018;20(17):11542–8.
87. Hossain SS, Bashkatov A, Yang X, Mutschke G, Eckert K. Force balance of hydrogen bubbles growing and oscillating on a microelectrode. Phys Rev E. 2022;106(3):1–15.
88. Babich A, Bashkatov A, Yang X, Mutschke G, Eckert K. In-situ measurements of temperature field and Marangoni convection at hydrogen bubbles using schlieren and PTV techniques. Int J Heat Mass Transf. 2023;215:124466.
89. Andersson M, Mularczyk A, Lamibrac A, Beale SB, Eller J, Lehnert W, et al. Modeling and synchrotron imaging of droplet detachment in gas channels of polymer electrolyte fuel cells. J Power Sources. 2018;404(August):159–71.
90. Jiao K, Li X. Water transport in polymer electrolyte membrane fuel cells. Prog Energy Combust Sci. 2011;37(3):221–91.
91. Pan M, Pan C, Li C, Zhao J. A review of membranes in proton exchange membrane fuel cells: Transport phenomena, performance and durability. Renew Sustain Energy Rev. 2021;141(January):110771.
92. Vetter R, Schumacher JO. Experimental parameter uncertainty in proton exchange membrane fuel cell modeling. Part I: Scatter in material parameterization. J Power Sources. 2019;438(August):227018.
93. Lamanna JM, Kandlikar SG. Determination of effective water vapor diffusion coefficient in pemfc gas diffusion layers. Int J Hydrogen Energy. 2011;36(8):5021–9.
94. Li YS, Zhao TS, Yang WW. Measurements of water uptake and transport properties in anion-exchange membranes. Int J Hydrogen Energy. 2010;35(11):5656–65.
95. Satjaritanun P, O'Brien M, Kulkarni D, Shimpalee S, Capuano C, Ayers KE, et al. Observation of Preferential Pathways for Oxygen Removal through Porous Transport Layers of Polymer Electrolyte Water Electrolyzers. iScience. 2020;23(12):101783.
96. De Angelis S, Schuler T, Charalambous MA, Marone F, Schmidt TJ, Büchi FN. Unraveling two-phase transport in porous transport layer materials for polymer electrolyte water electrolysis. J Mater Chem A. 2021;9(38):22102–13.
97. Lv H, Chen J, Zhou W, Shen X, Zhang C. Mechanism analyses and optimization strategies for performance improvement in low-temperature water electrolysis systems via the perspective of mass transfer: A review. Renew Sustain Energy Rev. 2023;183(June):113394.
98. Eriksson B, Santori PG, Lecoeur F, Dupont M, Jaouen F. Understanding the effects of operating conditions on the water management in anion exchange membrane fuel cells. J Power Sources. 2023;554(September 2022).
99. Hussaini IS, Wang CY. Visualization and quantification of cathode channel flooding in PEM fuel cells. J Power Sources. 2009;187(2):444–51.
100. Wang XR, Ma Y, Gao J, Li T, Jiang GZ, Sun ZY. Review on water management methods for proton

exchange membrane fuel cells. Int J Hydrogen Energy. 2021;46(22):12206–29.
101. Gutru R, Turtayeva Z, Xu F, Maranzana G, Vigolo B, Desforges A. A comprehensive review on water management strategies and developments in anion exchange membrane fuel cells. Int J Hydrogen Energy. 2020;45(38):19642–63.
102. Araújo F, Neto RC, Moita AS. Alkaline water electrolysis: Ultrasonic field and hydrogen bubble formation. Int J Hydrogen Energy. 2024;78(May):594–603.
103. Dickinson EJF, Wain AJ. The Butler-Volmer equation in electrochemical theory: Origins, value, and practical application. J Electroanal Chem. 2020;872:114145.
104. Vogt H. The Concentration Overpotential of Gas Evolving Electrodes as a Multiple Problem of Mass Transfer. J Electrochem Soc. 1990;137(4):1179–84.
105. Schmidt G, Suermann M, Bensmann B, Hanke-Rauschenbach R, Neuweiler I. Modeling Overpotentials Related to Mass Transport Through Porous Transport Layers of PEM Water Electrolysis Cells. J Electrochem Soc. 2020;167(11):114511.
106. Haverkort JW, Rajaei H. Electro-osmotic flow and the limiting current in alkaline water electrolysis. J Power Sources Adv. 2020;6(September):100034.
107. Wood DL, Borup RL. Estimation of Mass-Transport Overpotentials during Long-Term PEMFC Operation. J Electrochem Soc. 2010;157(8):B1251.
108. Iwata R, Zhang L, Wilke KL, Gong S, He M, Gallant BM, et al. Bubble growth and departure modes on wettable/non-wettable porous foams in alkaline water splitting. Joule. 2021;5(4):887–900.
109. Garcia-Navarro JC, Schulze M, Friedrich KA. Measuring and modeling mass transport losses in proton exchange membrane water electrolyzers using electrochemical impedance spectroscopy. J Power Sources. 2019;431(May):189–204.
110. Wu J, Yuan XZ, Wang H, Blanco M, Martin JJ, Zhang J. Diagnostic tools in PEM fuel cell research: Part I Electrochemical techniques. Int J Hydrogen Energy. 2008;33(6):1735–46.
111. Lazanas AC, Prodromidis MI. Electrochemical Impedance Spectroscopy─A Tutorial. ACS Meas Sci Au. 2023;3(3):162–93.
112. Rox H, Bashkatov A, Yang X, Loos S, Mutschke G, Gerbeth G, et al. Bubble size distribution and electrode coverage at porous nickel electrodes in a novel 3-electrode flow-through cell. Int J Hydrogen Energy. 2023;48(8):2892–905.
113. Krause L, Skibińska K, Rox H, Baumann R, Marzec MM, Yang X, et al. Hydrogen Bubble Size Distribution on Nanostructured Ni Surfaces: Electrochemically Active Surface Area Versus Wettability. ACS Appl Mater Interfaces. 2023;15(14):18290–9.
114. Kim BK, Kim MJ, Kim JJ. Impact of Surface Hydrophilicity on Electrochemical Water Splitting. ACS Appl Mater Interfaces. 2021;13(10):11940–7.
115. Li M, Li Y, Qin Y, Yin Y, Zhang J, Che Z. Water droplet detachment characteristics on surfaces of gas diffusion layers in PEMFCs. Int J Hydrogen Energy. 2022;47(18):10341–51.
116. Lee SK, Ito K. Cross-Sectional Visualization and Analysis of Droplet Behavior in Gas Flow Channel in PEFC. J Electrochem Soc. 2014;161(1):F58–66.
117. Hoeh MA, Arlt T, Manke I, Banhart J, Fritz DL, Maier W, et al. In operando synchrotron X-ray radiography studies of polymer electrolyte membrane water electrolyzers. Electrochem commun. 2015;55:55–9.
118. Ronovský M, Myllymäki M, Watier Y, Glatzel P, Strasser P, Bonastre AM, et al. Proton-exchange membrane fuel cell design for in-situ depth-sensitive X-ray absorption spectroscopy. J Power Sources. 2024;592(November 2023).
119. Kulkarni D, Huynh A, Satjaritanun P, O’Brien M, Shimpalee S, Parkinson D, et al. Elucidating effects of catalyst loadings and porous transport layer morphologies on operation of proton exchange membrane water electrolyzers. Appl Catal B Environ. 2022;308(February):121213.
120. Dedigama I, Angeli P, Van Dijk N, Millichamp J, Tsaoulidis D, Shearing PR, et al. Current density mapping and optical flow visualisation of a polymer electrolyte membrane water electrolyser. J Power Sources [Internet]. 2014;265:97–103. Available from: http://dx.doi.org/10.1016/j.jpowsour.2014.04.120
121. Esbo MR, Ranjbar AA, Rahgoshay SM. Analysis of water management in PEM fuel cell stack at dead-end mode using direct visualization. Renew Energy. 2020;162:212–21.
122. Wu TC, Djilali N. Experimental investigation of water droplet emergence in a model polymer electrolyte membrane fuel cell microchannel. J Power Sources. 2012;208:248–56.
123. Ibrahim-Rassoul N, Si-Ahmed EK, Serir A, Kessi A, Legrand J, Djilali N. Investigation of two-phase flow in a hydrophobic fuel-cell micro-channel. Energies. 2019;12(11).
124. Litster S, Sinton D, Djilali N. Ex situ visualization of liquid water transport in PEM fuel cell gas diffusion layers. J Power Sources. 2006;154(1):95–105.
125. Afra M, Nazari M, Kayhani MH, Sharifpur M, Meyer JP. 3D experimental visualization of water flooding in proton exchange membrane fuel cells. Energy. 2019;175:967–77.
126. Arbabi F, Kalantarian A, Abouatallah R, Wang R, Wallace JS, Bazylak A. Feasibility study of using

microfluidic platforms for visualizing bubble flows in electrolyzer gas diffusion layers. J Power Sources. 2014;258:142–9.
127. Bazylak A, Berejnov V, Markicevic B, Sinton D, Djilali N. Numerical and microfluidic pore networks: Towards designs for directed water transport in GDLs. Electrochim Acta. 2008;53(26):7630–7.
128. Lenormand R, Touboul E, Zarcone C. Numerical models and experiments on immiscible displacements in porous media. J Fluid Mech. 1988;189(1988):165–87.
129. Hughes RG, Blunt MJ. Pore scale modeling of rate effects in imbibition. Transp Porous Media. 2000;40(3):295–322.
130. Guo F, Aryana SA. An experimental investigation of flow regimes in imbibition and drainage using a microfluidic platform. Energies. 2019;12(7):1–13.
131. Hreiz R, Abdelouahed L, Fünfschilling D, Lapicque F. Electrogenerated bubbles induced convection in narrow vertical cells: PIV measurements and Euler-Lagrange CFD simulation. Chem Eng Sci. 2015;134:138–52.
132. Martin J, Oshkai P, Djilali N. Flow structures in a u-shaped fuel cell flow channel: Quantitative visualization using particle image velocimetry. J Fuel Cell Sci Technol. 2005;2(1):70–80.
133. Yoon SY, Ross JW, Mench MM, Sharp K V. Gas-phase particle image velocimetry (PIV) for application to the design of fuel cell reactant flow channels. J Power Sources. 2006;160(2 SPEC. ISS.):1017–25.
134. Han Y, Bashkatov A, Huang M, Eckert K, Mutschke G. Impact of tracer particles on the electrolytic growth of hydrogen bubbles. Phys Fluids. 2024;36(1).
135. Sharp K V., Yoon SY, Ross J, Mench M. Measurement of Fuel Cell Flowfields Using Particle Image Velocimetry. ECS Trans. 2006;1(6):571–80.
136. Hirono M, Suzuma I, Kurata T, Tadokoro R, Ejiri E, Takimoto M, et al. Measurements of Water Droplet Behavior in PEFC Channel Using PTV and High Accuracy Pressure Sensors. ECS Meet Abstr. 2009;MA2009-02(1):75–75.
137. Bilsing C, Janoske U, Czarske J, Büttner L, Burgmann S. 3D-3C measurements of flow reversal in small sessile drops in shear flow. Int J Multiph Flow. 2025;182(August 2024):105017.
138. Mei X, Yuan S, Zhao C, Yan X, Zhao CY, Wang Q. Measuring three-dimensional bubble dynamics for hydrogen production via water electrolysis. Phys Fluids. 2023;35(12).
139. Bilsing C, Radner H, Burgmann S, Czarske J, Büttner L. 3D Imaging with Double-Helix Point Spread Function and Dynamic Aberration Correction Using a Deformable Mirror. Opt Lasers Eng. 2022;154(February):107044.
140. Bilsing C, Nützenadel E, Burgmann S, Czarske J, Büttner L. Adaptive-optical 3D microscopy for microfluidic multiphase flows. Light Adv Manuf. 2024;5(0):1.
141. Bashkatov A, Bürkle F, Demirkır Ç, Ding W, Sanjay V, Babich A, et al. Electrolyte droplet spraying in H2 bubbles during water electrolysis under normal and microgravity conditions. Nat Commun . 2025;16(1).
142. Burgmann S, Krämer V, Dues M, Steinbock J, Büttner L, Czarske J, et al. Flow-measurements in the wake of an oscillating sessile droplet using laser-Doppler velocity profile sensor. Tech Mess. 2022;89(3):178–88.
143. Bürkle F, Moyon F, Feierabend L, Wartmann J, Heinzel A, Czarske J, et al. Investigation and equalisation of the flow distribution in a fuel cell stack. J Power Sources. 2020;448(December 2019):227546.
144. Bürkle F, Czarske J, Büttner L. Simultaneous velocity profile and temperature profile measurements in microfluidics. Flow Meas Instrum. 2022;83(January):102106.
145. Bashkatov A, Babich A, Hossain SS, Yang X, Mutschke G, Eckert K. H 2 bubble motion reversals during water electrolysis. J Fluid Mech. 2023;958:1–14.
146. Yeganeh M, Cheng Q, Dharamsi A, Karimkashi S, Kuusela-Opas J, Kaario O, et al. Visualization and comparison of methane and hydrogen jet dynamics using schlieren imaging. Fuel. 2023;331(P1):125762.
147. Babich A, Bashkatov A, Eftekhari M, Yang X, Strasser P, Mutschke G, et al. Oxygen versus Hydrogen Bubble Dynamics during Water Electrolysis at Microelectrodes. Phys Rev Appl. 2025;10(1):1.
148. Babich A, Mutschke G, Bashkatov A, Rox H, Eftekhari M, Yang X, et al. Solutal Marangoni convection at growing oxygen bubbles during water electrolysis. Phys Rev Res. 2025;7(2):23189.
149. Lee J, Kim Y, Kim W, Lee K. Visualization of Injected Fuel Vaporization Using Background-Oriented Schlieren Method. Energies. 2024;17(19).
150. Schmieder F, Kinaci ME, Wartmann J, König J, Büttner L, Czarske J, et al. Investigation of the flow field inside the manifold of a real operated fuel cell stack using optical measurements and Computational Fluid Mechanics. J Power Sources. 2016;304:155–63.
151. Devi N, Ray S, Shukla A, Bhat SD, Pesala B. Non-invasive macroscopic and molecular quantification of water in Nafion® and SPEEK Proton Exchange Membranes using terahertz spectroscopy. J Memb Sci. 2019;588(March):117183.
152. Alves-Lima DF, Williams BM, Schlegl H, Gupta G, Letizia R, Dawson R, et al. Visualizing water inside an operating proton exchange membrane fuel cell with video-rate terahertz imaging. Fuel Cells. 2022;22(6):229–40.

153. Alves-Lima DF, Li X, Coulson B, Nesling E, Ludlam GAH, Degl'Innocenti R, et al. Evaluation of water states in thin proton exchange membrane manufacturing using terahertz time-domain spectroscopy. J Memb Sci. 2022;647(February):1–11.
154. Yurchenko SO, Zaytsev KI. Spectroscopy of nafion in terahertz frequency range. J Appl Phys. 2014;116(11).
155. Haidekker MA. Medical Imaging Technology. In: SpringerBriefs in Physics. 2013. p. 1–125.
156. Neutron Physics [Internet]. Available from: https://www.psi.ch/en/niag/neutron-physics
157. J. H. Hubbell, S. M. Seltzer. NIST Standard Reference Database 126. 2004. X-Ray Mass Attenuation Coefficients. Available from: https://www.nist.gov/pml/x-ray-mass-attenuation-coefficients
158. Kardjilov N, Fiori F, Giunta G, Hilger A, Rustichelli F, Strobl M, et al. Neutron tomography for archaeological investigations. J Neutron Res. 2006;14(1):29–36.
159. Altus SJ, Inkson BJ, Hack J. Complementary X-ray and neutron imaging of water electrolysers for green hydrogen production. J Mater Chem A. 2024;
160. Maier M, Dodwell J, Ziesche R, Tan C, Heenan T, Majasan J, et al. Mass transport in polymer electrolyte membrane water electrolyser liquid-gas diffusion layers: A combined neutron imaging and X-ray computed tomography study. J Power Sources. 2020;455.
161. Ziesche RF, Hack J, Rasha L, Maier M, Tan C, Heenan TMM, et al. High-speed 4D neutron computed tomography for quantifying water dynamics in polymer electrolyte fuel cells. Nat Commun. 2022;13(1):1–11.
162. Leonard E, Shum AD, Danilovic N, Capuano C, Ayers KE, Pant LM, et al. Interfacial analysis of a PEM electrolyzer using X-ray computed tomography. Sustain Energy Fuels. 2020;4(2):921–31.
163. Koch S, Disch J, Kilian SK, Han Y, Metzler L, Tengattini A, et al. Water management in anion-exchange membrane water electrolyzers under dry cathode operation. RSC Adv. 2022;12(32):20778–84.
164. Manke I, Hartnig C, Grünerbel M, Lehnert W, Kardjilov N, Haibel A, et al. Investigation of water evolution and transport in fuel cells with high resolution synchrotron x-ray radiography. Appl Phys Lett. 2007;90(17).
165. Oksuz I, Bisbee M, Hall J, Cherepy N, Cao L. Quantifying spatial resolution in a fast neutron radiography system. Nucl Instruments Methods Phys Res Sect A Accel Spectrometers, Detect Assoc Equip. 2022;1027(September 2021):166331.
166. Chen YS, Peng H, Hussey DS, Jacobson DL, Tran DT, Abdel-Baset T, et al. Water distribution measurement for a PEMFC through neutron radiography. J Power Sources. 2007;170(2):376–86.
167. Trabold TA, Owejan JP, Jacobson DL, Arif M, Huffman PR. In situ investigation of water transport in an operating PEM fuel cell using neutron radiography: Part 1 - Experimental method and serpentine flow field results. Int J Heat Mass Transf. 2006;49(25–26):4712–20.
168. Zlobinski M, Schuler T, Büchi FN, Schmidt TJ, Boillat P. Transient and Steady State Two-Phase Flow in Anodic Porous Transport Layer of Proton Exchange Membrane Water Electrolyzer. J Electrochem Soc. 2020;167(8):084509.
169. Renz S, Arlt T, Kardjilov N, Helfen L, Couture C, Tengattini A, et al. Operando Neutron Radiography Measurements of a Zero-Gap Alkaline Electrolysis Cell. ECS Meet Abstr. 2022;MA2022-02(39):1448–1448.
170. Xu L, Trogadas P, Zhou S, Jiang S, Wu Y, Rasha L, et al. A Scalable and Robust Water Management Strategy for PEMFCs: Operando Electrothermal Mapping and Neutron Imaging Study. Adv Sci. 2024;2404350:1–12.
171. Cooper NJ, Santamaria AD, Becton MK, Park JW. Neutron radiography measurements of in-situ PEMFC liquid water saturation in 2D & 3D morphology gas diffusion layers. Int J Hydrogen Energy. 2017;42(25):16269–78.
172. Martinez N, Peng Z, Morin A, Porcar L, Gebel G, Lyonnard S. Real time monitoring of water distribution in an operando fuel cell during transient states. J Power Sources. 2017;365:230–4.
173. Owejan JP, Trabold TA, Jacobson DL, Baker DR, Hussey DS, Arif M. In situ investigation of water transport in an operating PEM fuel cell using neutron radiography: Part 2 - Transient water accumulation in an interdigitated cathode flow field. Int J Heat Mass Transf. 2006;49(25–26):4721–31.
174. Chen Y, Stelmacovich G, Mularczyk A, Parkinson D, Babu SK, Forner-Cuenca A, et al. A Viewpoint on X-ray Tomography Imaging in Electrocatalysis. ACS Catal. 2023;13(15):10010–25.
175. Hack J, Rasha L, Cullen PL, Bailey JJ, Neville TP, Shearing PR, et al. Use of X-ray computed tomography for understanding localised, along-the-channel degradation of polymer electrolyte fuel cells. Electrochim Acta [Internet]. 2020;352:136464. Available from: https://doi.org/10.1016/j.electacta.2020.136464
176. White RT, Wu A, Najm M, Orfino FP, Dutta M, Kjeang E. 4D in situ visualization of electrode morphology changes during accelerated degradation in fuel cells by X-ray computed tomography. J Power Sources. 2017;350:94–102.
177. Xu H, Bührer M, Marone F, Schmidt TJ, Büchi FN, Eller J. Fighting the Noise: Towards the Limits of Subsecond X-ray Tomographic Microscopy of PEFC. ECS Meet Abstr. 2017;MA2017-02(32):1429.

178. Leonard E, Shum AD, Normile S, Sabarirajan DC, Yared DG, Xiao X, et al. Operando X-ray tomography and sub-second radiography for characterizing transport in polymer electrolyte membrane electrolyzer. Electrochim Acta. 2018;276:424–33.
179. Rahimian P, Battrell L, Anderson R, Zhu N, Johnson E, Zhang L. Investigation of time dependent water droplet dynamics on porous fuel cell material via synchrotron based X-ray imaging technique. Exp Therm Fluid Sci. 2018;97(August 2017):237–45.
180. Liu J, Talarposhti MR, Asset T, Sabarirajan DC, Parkinson DY, Atanassov P, et al. Understanding the Role of Interfaces for Water Management in Platinum Group Metal-Free Electrodes in Polymer Electrolyte Fuel Cells. ACS Appl Energy Mater. 2019;2(5):3542–53.
181. Nagai Y, Eller J, Hatanaka T, Yamaguchi S, Kato S, Kato A, et al. Improving water management in fuel cells through microporous layer modifications: Fast operando tomographic imaging of liquid water. J Power Sources. 2019;435(March):226809.
182. Eller J, Roth J, Marone F, Stampanoni M, Büchi FN. Operando Properties of Gas Diffusion Layers: Saturation and Liquid Permeability. J Electrochem Soc. 2017;164(2):F115–26.
183. Xu H, Nagashima S, Nguyen HP, Kishita K, Marone F, Büchi FN, et al. Temperature dependent water transport mechanism in gas diffusion layers revealed by subsecond operando X-ray tomographic microscopy. J Power Sources. 2021;490(December 2020):229492.
184. Langel W. Introduction to neutron scattering [Internet]. Vol. 9, ChemTexts. Springer International Publishing; 2023. 1–55 p.
185. Mosdale R, Gebel G, Pineri M. Water profile determination in a running proton exchange membrane fuel cell using small-angle neutron scattering. J Memb Sci. 1996;118(2):269–77.
186. Magnussen OM, Drnec J, Qiu C, Martens I, Huang JJ, Chattot R, et al. In Situ and Operando X-ray Scattering Methods in Electrochemistry and Electrocatalysis. Chem Rev. 2024;124(3):629–721.
187. Lee J, Escribano S, Micoud F, Gebel G, Lyonnard S, Porcar L, et al. In Situ Measurement of Ionomer Water Content and Liquid Water Saturation in Fuel Cell Catalyst Layers by High-Resolution Small-Angle Neutron Scattering. ACS Appl Energy Mater. 2020;3(9):8393–401.
188. Martens I, Vamvakeros A, Chattot R, Blanco M V., Rasola M, Pusa J, et al. X-ray transparent proton-exchange membrane fuel cell design for in situ wide and small angle scattering tomography. J Power Sources. 2019;437(July).
189. Aliyah K, Appel C, Lazaridis T, Prehal C, Ammann M, Xu L, et al. Operando Scanning Small-/Wide-Angle X-ray Scattering for Polymer Electrolyte Fuel Cells: Investigation of Catalyst Layer Saturation and Membrane Hydration- Capabilities and Challenges. ACS Appl Mater Interfaces. 2024;16(20):25938–52.
190. Selamet OF, Pasaogullari U, Spernjak D, Hussey DS, Jacobson DL, Mat M. In Situ Two-Phase Flow Investigation of Proton Exchange Membrane (PEM) Electrolyzer by Simultaneous Optical and Neutron Imaging. ECS Meet Abstr. 2011;MA2011-02(16):978–978.
191. Eller J, Roth J, Marone F, Stampanoni M, Wokaun A, Büchi FN. Implications of polymer electrolyte fuel cell exposure to synchrotron radiation on gas diffusion layer water distribution. J Power Sources. 2014;245:796–800.
192. Lee CY, Chen CH, Chuang SM, Dai CL, Lai BJ, Chen SY, et al. Real-time data acquisition inside high-pressure PEM water electrolyzer. Sensors Actuators A Phys. 2024;372(April):115318.
193. Kim T, Kim Y, Han J, Yu S. Multiple in-situ measurement of water transport in the bipolar plate of proton exchange membrane fuel cell. Int J Heat Mass Transf. 2024;225(March):125269.
194. David N, Von Schilling K, Wild PM, Djilali N. In situ measurement of relative humidity in a PEM fuel cell using fibre Bragg grating sensors. Int J Hydrogen Energy. 2014;39(31):17638–44.
195. Zhao J, Tu Z, Chan SH. In-situ measurement of humidity distribution and its effect on the performance of a proton exchange membrane fuel cell. Energy. 2022;239:122270.
196. Shao H, Qiu D, Peng L, Yi P, Lai X. In-situ measurement of temperature and humidity distribution in gas channels for commercial-size proton exchange membrane fuel cells. J Power Sources. 2019;412(July 2018):717–24.
197. Zhou Z, Ye L, Qiu D, Peng L, Lai X. Experimental investigation and decoupling of voltage losses distribution in proton exchange membrane fuel cells with a large active area. Chem Eng J. 2023;452(P4):139497.
198. David NA, Wild PM, Jensen J, Navessin T, Djilali N. Simultaneous In Situ Measurement of Temperature and Relative Humidity in a PEMFC Using Optical Fiber Sensors. J Electrochem Soc. 2010;157(8):B1173.
199. Lee CY, Hsieh WJ, Wu GW. Embedded flexible micro-sensors in MEA for measuring temperature and humidity in a micro-fuel cell. J Power Sources. 2008;181(2):237–43.
200. Tsujikawa J, Minami R, Araki T. In Situ Humidity Measurements at the CL Surface By MEMS-Based Sensors. ECS Meet Abstr. 2015;MA2015-02(37):1475–1475.
201. Lee CY, Chen CH, Jung G Bin, Zheng YX, Liu YC. PEMWE with internal real-time microscopic monitoring function. Membranes (Basel). 2021;11(2).
202. Lee CY, Chen CH, Chuang HC, Hsieh H Te, Chiu YC. Long-Acting Real-Time Microscopic Monitoring

Inside the Proton Exchange Membrane Water Electrolyzer. Sensors. 2023;23(12).
203. Hauer KH, Potthast R, Wüster T, Stolten D. Magnetotomography - A new method for analysing fuel cell performance and quality. J Power Sources. 2005;143(1–2):67–74.
204. Sun Y, Mao L, Hu Z, Zhang X, Peng R. Magnetic Array-Aided Visualizing PEMFC Degradation Heterogeneity. Advanced Science. 2024.
205. Sun Y, Mao L, He K, Liu Z, Lu S. Imaging PEMFC performance heterogeneity by sensing external magnetic field. Cell Reports Phys Sci. 2022;3(10):101083.
206. Akimoto Y, Shibata M, Tsuzuki Y, Okajima K, Suzuki S nosuke. In-situ on-board evaluation and control of proton exchange membrane fuel cells using magnetic sensors. Appl Energy. 2023;351(May):121873.
207. David N, Djilali N, Wild P. Fiber Bragg grating sensor for two-phase flow in microchannels. Microfluid Nanofluidics. 2012;13(1):99–106.
208. Wang H, Morando S, Gaillard A, Hissel D. Sensor development and optimization for a proton exchange membrane fuel cell system in automotive applications. J Power Sources. 2021;487(January):229415.
209. Majasan JO, Robinson JB, Owen RE, Maier M, Radhakrishnan ANP, Pham M, et al. Recent advances in acoustic diagnostics for electrochemical power systems. JPhys Energy. 2021;3(3).
210. Minnaert M. XVI. On musical air-bubbles and the sounds of running water . London, Edinburgh, Dublin Philos Mag J Sci. 1933;16(104):235–48.
211. Leighton TG, Apfel RE. The Acoustic Bubble. Vol. 96, The Journal of the Acoustical Society of America. 1994. 2616–2616 p.
212. Hassan F, Mahmood AK Bin, Yahya N, Saboor A, Abbas MZ, Khan Z, et al. State-of-the-Art Review on the Acoustic Emission Source Localization Techniques. IEEE Access. 2021;9:101246–66.
213. Maier M, Meyer Q, Majasan J, Owen RE, Robinson JB, Dodwell J, et al. Diagnosing Stagnant Gas Bubbles in a Polymer Electrolyte Membrane Water Electrolyser Using Acoustic Emission. Front Energy Res. 2020;8(October):1–6.
214. Maier M, Meyer Q, Majasan J, Tan C, Dedigama I, Robinson J, et al. Operando flow regime diagnosis using acoustic emission in a polymer electrolyte membrane water electrolyser. J Power Sources. 2019;424(March):138–49.
215. Bethapudi VS, Hack J, Trogadas P, Hinds G, Shearing PR, Brett DJL, et al. Hydration state diagnosis in fractal flow-field based polymer electrolyte membrane fuel cells using acoustic emission analysis. Energy Convers Manag. 2020;220(April):113083.
216. Al-Rweg M, Ahmeda K, Albarbar A. Acoustical Characteristics of Proton Exchange Membrane Fuel Cells. IEEE Access. 2021;9:81068–77.
217. Legros B, Thivel PX, Bultel Y, Boinet M, Nogueira RP. Acoustic emission: Towards a real-time diagnosis technique for Proton exchange membrane fuel cell operation. J Power Sources. 2010;195(24):8124–33.
218. Murakami K, Yamakawa Y, Zhao J, Johnsen E, Ando K. Ultrasound-induced nonlinear oscillations of a spherical bubble in a gelatin gel. J Fluid Mech. 2021;924:1–17.
219. Dou Z, Rox H, Ramos Z, Baumann R, Ravishankar R, Czurratis P, et al. Scanning Acoustic Microscopy for Quantifying Bubble Evolution in Alkaline Water Electrolyzers. J Power Sources. 2025; 660: 238575.
220. Solids and Metals - Speed of Sound [Internet]. Available from: https://www.engineeringtoolbox.com/sound-speed-solids-d_713.html
221. Merabet NH, Kerboua K. Green hydrogen from sono-electrolysis: A coupled numerical and experimental study of the ultrasound assisted membraneless electrolysis of water supplied by PV. Fuel. 2024;356(July 2023):129625.
222. Sharifishourabi M, Dincer I, Mohany A. Investigation and assessment of newly developed sonochemical and sonoelectrochemical systems. Int J Hydrogen Energy. 2025;163(August):150694.
223. Gravelle J, Avramovic V, Hallez L, Hihn JY, Pollet BG. Effects of a perpendicular ultrasonic field on planar and porous electrodes for hydrogen production in alkaline conditions. Ultrason Sonochem. 2025;120(July).
224. Quarato CMI, Lacedonia D, Salvemini M, Tuccari G, Mastrodonato G, Villani R, et al. A Review on Biological Effects of Ultrasounds: Key Messages for Clinicians. Diagnostics. 2023;13(5):1–29.
225. Maier M, Owen RE, Pham MTM, Dodwell J, Majasan J, Robinson JB, et al. Acoustic time-of-flight imaging of polymer electrolyte membrane water electrolysers to probe internal structure and flow characteristics. Int J Hydrogen Energy. 2021;46(21):11523–35.
226. Dou Z, Tropf L, Hoster H, Schmidt H, Czarske J, Weik D. Advanced Ultrasonic Diagnostic Technology Towards Green Hydrogen Energy Systems. In: IEEE International Ultrasonics Symposium, IUS. IEEE; 2023. p. 1–4.
227. Weik D, Grüter L, Räbiger D, Singh S, Vogt T, Eckert S, et al. Ultrasound Localization Microscopy in Liquid Metal Flows. Appl Sci. 2022;12(9).
228. Weik D, Dou Z, Räbiger D, Vogt T, Eckert S, Czarske J, et al. Uncertainty Quantification of Super-Resolution Flow Mapping in Liquid Metals using Ultrasound Localization Microscopy. 2024;(April). Available from: http://arxiv.org/abs/2404.10840

229. Yu J, Lavery L, Kim K. Super-resolution ultrasound imaging method for microvasculature in vivo with a high temporal accuracy. Sci Rep. 2018;(August):1–11.
230. Kupsch C, Feierabend L, Nauber R, Buttner L, Czarske J. Ultrasound Super-Resolution Flow Measurement of Suspensions in Narrow Channels. IEEE Trans Ultrason Ferroelectr Freq Control. 2021;68(3):807–17.
231. Dou Z, Xu Y, Wei Z, Emmerich H, Czarske J, Weik D. Towards Monitoring Water Content in Membrane Electrode Assembly of Low Temperature Fuel Cells Using Ultrasound. In: 2024 IEEE Ultrasonics, Ferroelectrics, and Frequency Control Joint Symposium (UFFC-JS). IEEE; 2024. p. 1–4.
232. Huang M, Kirkaldy N, Zhao Y, Patel Y, Cegla F, Lan B. Quantitative characterisation of the layered structure within lithium-ion batteries using ultrasonic resonance. J Energy Storage. 2022;50(April):104585.
233. Sablowski J, Zhao Z, Kupsch C. Ultrasonic Guided Waves for Liquid Water Localization in Fuel Cells: An Ex Situ Proof of Principle. Sensors. 2022;22(21).
234. Sablowski J, Lugovtsova Y, Bulling J, Kupsch C. Experimental Study of Ultrasonic Guided Waves in a Bipolar Plate of a Fuel Cell. IEEE Int Ultrason Symp IUS. 2023;1–4.
235. Dou Z, Fang B, Tropf L, Hoster H, Schmidt H, Czarske J, et al. A Water Monitoring System for Proton Exchange Membrane Fuel Cells Based on Ultrasonic Lamb Waves: An Ex Situ Proof of Concept. IEEE Trans Instrum Meas. 2023;72:1–12.
236. Heckel R. Deep Learning for Computational Imaging. Adv Photonics. 2025;7(5):1–219.